\documentclass[superscriptaddress,twocolumn,showpacs,a4paper,longbibliography,
amssymb,amsmath,nobibnotes,aps,prd,
showkeys,
nofootinbib,notitlepage]{revtex4-1}
\usepackage{verbatim}
\usepackage[T1]{fontenc}
\usepackage[utf8]{inputenc}
\usepackage[american]{babel}
\usepackage{epsfig}
\usepackage{booktabs}
\usepackage{dcolumn}
\usepackage{amsmath}
\usepackage{mathtools}
\usepackage{subcaption}
\usepackage{amsfonts}
\usepackage{amssymb}
\usepackage{ulem}
\usepackage{epstopdf}
\usepackage{bm}
\usepackage{siunitx}
\usepackage{braket}
\usepackage{enumitem}
\usepackage{soul}
\usepackage[table]{xcolor}
\usepackage{color}
\usepackage{transparent}
\usepackage{pifont}
\usepackage{enumitem}

\definecolor{navyblue}{rgb}{0.0, 0.0, 0.5}
\definecolor{royalblue}{rgb}{0.25, 0.41, 0.88}
\definecolor{cadmiumgreen}{rgb}{0.0, 0.42, 0.24}
\definecolor{blue-violet}{rgb}{0.54, 0.17, 0.89}
\definecolor{darkviolet}{rgb}{0.58, 0.0, 0.83}
\definecolor{orange(colorwheel)}{rgb}{1.0, 0.5, 0.0}

\usepackage{hyperref}
\hypersetup{
    colorlinks=true, 
    linkcolor=royalblue, 
    citecolor=blue}

\usepackage{booktabs}
\usepackage{multirow}
\usepackage{dcolumn}
\usepackage{colortbl}

\begin{document}

\title{Beyond Two Parameters: Revisiting Dark Energy with the Latest Cosmic Probes}

\author{Hanyu Cheng}
\email{hcheng19@sheffield.ac.uk}
\affiliation{Tsung-Dao Lee Institute (TDLI),
No.\ 1 Lisuo Road, 
 201210 Shanghai, China}
\affiliation{School of Physics and Astronomy, Shanghai Jiao Tong University,
800 Dongchuan Road, 200240 Shanghai, China}
\affiliation{School of Mathematical and Physical Sciences, University of Sheffield, Hounsfield Road, Sheffield S3 7RH, United Kingdom}

\author{Supriya Pan}
\email{supriya.maths@presiuniv.ac.in}
\affiliation{Department of Mathematics, Presidency University, 86/1 College Street,  Kolkata 700073, India}
\affiliation{Institute of Systems Science, 
Durban University of Technology, Durban 4000, Republic of South Africa}

\author{Eleonora Di Valentino}
\email{e.divalentino@sheffield.ac.uk}
\affiliation{School of Mathematical and Physical Sciences, University of Sheffield, Hounsfield Road, Sheffield S3 7RH, United Kingdom}

\begin{abstract}
Dark energy (DE) models with many free parameters are often considered excessive, as constraining all parameters poses a significant challenge. While such models offer greater flexibility to probe the DE sector in more detail. With the rapid advancement of astronomical surveys and the availability of diverse datasets,
it is timely to examine whether current combined observations can effectively constrain an extended parameter space in DE models.
This article investigates a four-parameter dynamical DE  model that spans a broad region of the universe's expansion history through four key parameters: present-day value of the DE equation of state ($w_0$), its initial value ($w_m$), scale factor depicting  transition from $w_m$ to $w_0$  ($a_t$), and  steepness of this transition ($\Delta_{\rm de}$). We constrain the model using cosmic microwave background data from Planck, BAO from DESI DR2, and three distinct compilations of Type Ia Supernovae: PantheonPlus, DESY5, and Union3. Our results show that constraining all four parameters remains challenging:
$a_t$ is not constrained by any dataset, 
constraints on $w_m$ and
$\Delta_{\rm de}$ remain weak,
only $w_0$ is well
constrained across all datasets. The results further show that $w_0 > -1$, while $w_m$ is negative, indicating a phantom-like behaviour at early times. 
Interestingly, despite its larger parameter space, the proposed  model shows
a preference over the $\Lambda$CDM and $w_0w_a$CDM scenarios for certain combined
datasets, according to both $\Delta \chi^2$ and Bayesian evidence, although this
preference is not strong. 

\end{abstract}

\maketitle

\section{Introduction}
\label{sec:Introduction}

The late-time accelerating expansion of our universe~\cite{SupernovaSearchTeam:1998fmf, SupernovaCosmologyProject:1998vns} was a groundbreaking discovery that opened new avenues in cosmology and astrophysics. This observation clearly indicated that a cosmological model based solely on ordinary matter cannot explain the observed acceleration. Instead, it requires the presence of a hypothetical component with sufficiently negative pressure, commonly referred to as dark energy (DE). However, the true nature, origin, and evolution of this mysterious component remain elusive.
The simplest and most widely studied candidate for DE is a positive cosmological constant $\Lambda$, corresponding to the vacuum energy within the framework of Einstein's General Relativity (GR). When included in the gravitational field equations of GR, $\Lambda$ alone can account for the accelerating expansion. It is characterized by a barotropic equation of state $w_{\Lambda} = p_{\Lambda}/\rho_{\Lambda} = -1$, where $\rho_{\Lambda}$ and $p_{\Lambda}$ are the energy density and pressure of the vacuum, respectively.\footnote{As $\Lambda > 0$, vacuum energy density is positive, which implies a negative pressure and thus drives acceleration.}
Current cosmological observations suggest that $\Lambda$ contributes roughly 68\% of the total energy budget of the universe. The remaining 32\% is dominated by cold (pressureless) dark matter (CDM), accounting for about 28\%. This concordance model, composed of $\Lambda$ and CDM, is known as the $\Lambda$CDM model. It has become the standard cosmological paradigm due to its success in fitting a wide range of independent observations.
Nonetheless, it is well known that the $\Lambda$CDM model faces several unresolved challenges, such as the cosmological constant problem~\cite{Weinberg:1988cp}, the cosmic coincidence problem~\cite{Zlatev:1998tr}, and a series of persistent cosmological tensions~\cite{DiValentino:2021izs,Perivolaropoulos:2021jda,Abdalla:2022yfr,CosmoVerseNetwork:2025alb}. These issues suggest that $\Lambda$CDM may be an incomplete description of the universe’s evolution. As a result, numerous efforts have been made to extend or revise the standard model, either by introducing time-dependent dark energy within GR, modifying gravity, or proposing entirely new frameworks.
Many such models have been proposed to better capture the universe's expansion history (see, for example, the reviews~\cite{Peebles:2002gy,Nojiri:2006ri,Copeland:2006wr,Sahni:2006pa,Sotiriou:2008rp,DeFelice:2010aj,Clifton:2011jh,Bamba:2012cp,Koyama:2015vza,Cai:2015emx,Bahamonde:2021gfp} and references therein). However, none of these alternatives has yet emerged as a definitive model capable of explaining all current observations and resolving the existing anomalies. This motivates the search for new cosmological models that can better address the shortcomings of $\Lambda$CDM and reconcile the discrepancies in current astronomical data.

One of the simplest yet compelling extensions of the $\Lambda$CDM model involves modifying the DE sector through its barotropic equation of state (EoS), $w_{\rm de} = p_{\rm de}/\rho_{\rm de}$, where $p_{\rm de}$ and $\rho_{\rm de}$ are the pressure and energy density of DE, respectively. This EoS can be either constant—departing from $w = -1$, or time-dependent. A constant EoS with $w_{\rm de} \neq -1$ gives rise to the well-known $w$CDM cosmological model, in which DE is confined to either the quintessence regime ($w_{\rm de} > -1$) or the phantom regime ($w_{\rm de} < -1$), but disallows any crossing of the cosmological constant boundary, $w_{\rm de} = -1$.
By contrast, time-dependent formulations of $w_{\rm de}$ are more flexible and allow for richer phenomenology. These models are particularly intriguing in light of recent observations from DESI, which do not rule out a dynamical DE EoS~\cite{DESI:2024mwx,DESI:2025zgx}. DESI’s analysis, based on the widely-used Chevallier–Polarski–Linder (CPL) parametrization ($w_{\rm de} (a) = w_0 + w_a (1 - a)$)~\cite{Chevallier:2000qy,Linder:2002et}, suggests that models with evolving EoS remain viable~\cite{DESI:2024mwx,DESI:2025zgx}. Similar conclusions were also reported in other recent analyses considering different DE parametrizations~\cite{Giare:2024gpk,Wolf:2025jlc}.
Although the choice of $w_{\rm de}(a)$ is often phenomenological due to the lack of a guiding theoretical principle, this freedom has led to a wide range of models with varying numbers of free parameters: from zero-parameter forms to highly flexible multi-parameter scenarios~\cite{Cooray:1999da,Efstathiou:1999tm,Chevallier:2000qy,Linder:2002et,Corasaniti:2002vg,Bassett:2002fe,Kunz:2003iz,Corasaniti:2004sz,Wetterich:2004pv,Alam:2004jy,Linder:2005ne,Gong:2005de,Jassal:2005qc,Feng:2004ff,Melchiorri:2006jy,Zhang:2006em,Barboza:2008rh,Ma:2011nc,Sendra:2011pt,DeFelice:2012vd,Li:2012vn,Feng:2012gf,Novosyadlyj:2013nya,Akarsu:2015yea,Dimakis:2016mip,Yang:2017alx,Rezaei:2017yyj,Pan:2017zoh,Wang:2017lai,Yang:2018prh,Yang:2018qmz,Rezaei:2019hvb,Li:2019yem,Pan:2019hac,Pan:2019brc,Benaoum:2020qsi,Yang:2021eud,Yang:2021flj,Sharma:2022ifr,vonMarttens:2022xyr,Yao:2022jrw,Rezaei:2023xkj,Rezaei:2024vtg,Reyhani:2024cnr,Najafi:2024qzm,Giare:2024gpk,Giare:2024ocw,RoyChoudhury:2024wri,Escamilla:2024fzq,Gao:2024ily,Giare:2025pzu,Wolf:2025jlc,Paliathanasis:2025cuc,Liu:2025mub,Kessler:2025kju,Santos:2025wiv,Scherer:2025esj,RoyChoudhury:2025dhe,Cheng:2025hug,Sabogal:2025jbo,Herold:2025hkb,Lee:2025pzo,Silva:2025twg,Ishak:2025cay,Cortes:2024lgw,Shlivko:2024llw,Luongo:2024fww,Yin:2024hba,Gialamas:2024lyw,Dinda:2024kjf,Wang:2024dka,Ye:2024ywg,Tada:2024znt,Carloni:2024zpl,Chan-GyungPark:2024mlx,DESI:2024kob,Ramadan:2024kmn,Notari:2024rti,Orchard:2024bve,Hernandez-Almada:2024ost,Malekjani:2024bgi,Reboucas:2024smm,Park:2024pew,Menci:2024hop,Li:2024qus,Li:2024bwr,Notari:2024zmi,Fikri:2024klc,Jiang:2024xnu,Zheng:2024qzi,Gomez-Valent:2024ejh,Lewis:2024cqj,Du:2024pai,Shajib:2025tpd,Chaussidon:2025npr,Pang:2025lvh,Teixeira:2025czm,Specogna:2025guo,Cheng:2025lod,Ozulker:2025ehg,Gonzalez-Fuentes:2025lei,Song:2025bio,Li:2025vuh,Rezaei:2025vhb,Paliathanasis:2025mvy,Nair:2025uyn,Zhou:2025ugf}.
Despite this proliferation, no single parametrization has emerged as strongly preferred by data. In this context, any $w_{\rm de}$ model that remains consistent with observations continues to be worth investigating, especially as upcoming high-precision surveys may help distinguish among them or reveal subtle signatures of dark energy evolution.

Given the absence of theoretical guidance on the optimal number of free parameters in a $w_{\rm de}$ model, and the limited exploration in the literature of DE parametrizations with more than two parameters, in this article we investigate a four-parameter $w_{\rm de}$ model originally proposed in~\cite{Corasaniti:2002vg}. The novelty of this parametrization lies in its ability to simultaneously capture key aspects of DE dynamics: the EoS at both early and late times, the scale factor at which the transition between these regimes occurs, and the sharpness of this transition.
To the best of our knowledge, this model has been seldom studied in the context of DE cosmology. One likely reason is its relatively high computational cost due to the increased number of free parameters, which may lead to degeneracies and weaker constraints. However, with the rapid improvement in the precision of cosmological observations and the availability of extensive datasets spanning a wide redshift range, a re-examination of such a flexible model is timely and potentially insightful for uncovering subtle features of DE.

The structure of the paper is as follows. In Section~\ref{sec-2}, we introduce the dynamical dark energy (DDE) model under consideration. Section~\ref{sec-data} describes the observational datasets used to constrain the model. In Section~\ref{sec-results}, we present and analyze the results. Finally, Section~\ref{sec-summary} provides a summary and concluding remarks.

\section{Model}
\label{sec-2}

\begin{figure}[htbp]
    \centering
    \includegraphics[width=1.0\linewidth]{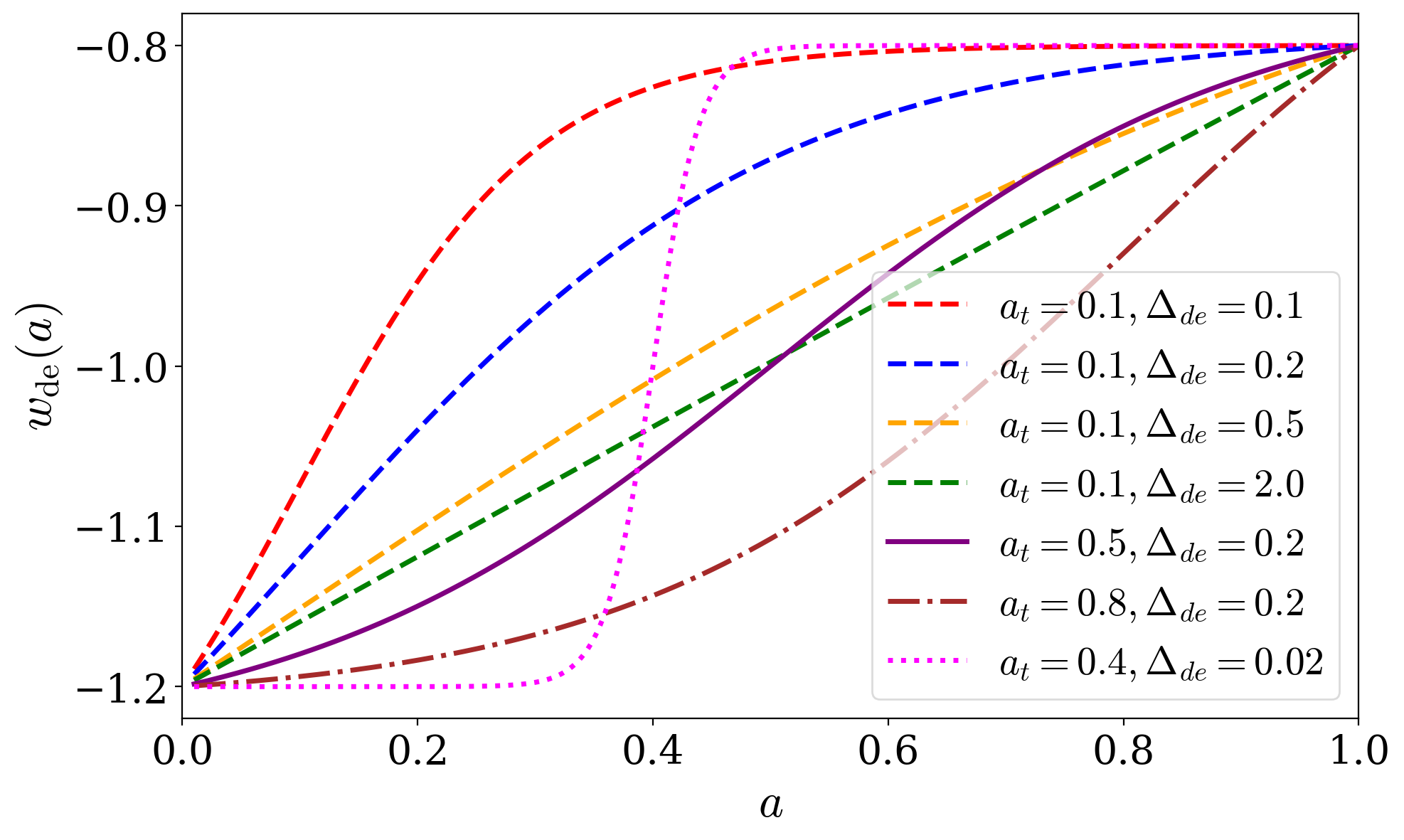}
    \caption{Evolution of $w_{\mathrm{de}}(a)$ for different sets of values of $a_{\mathrm{t}}$ and $\Delta_{\mathrm{de}}$, with the parameters $w_0$ and $w_{\mathrm{m}}$ fixed to $-0.8$ and $-1.2$, respectively.}
    \label{fig:w_de}
\end{figure}

\begin{figure}[htbp]
    \centering
    \includegraphics[width=1.0\linewidth]{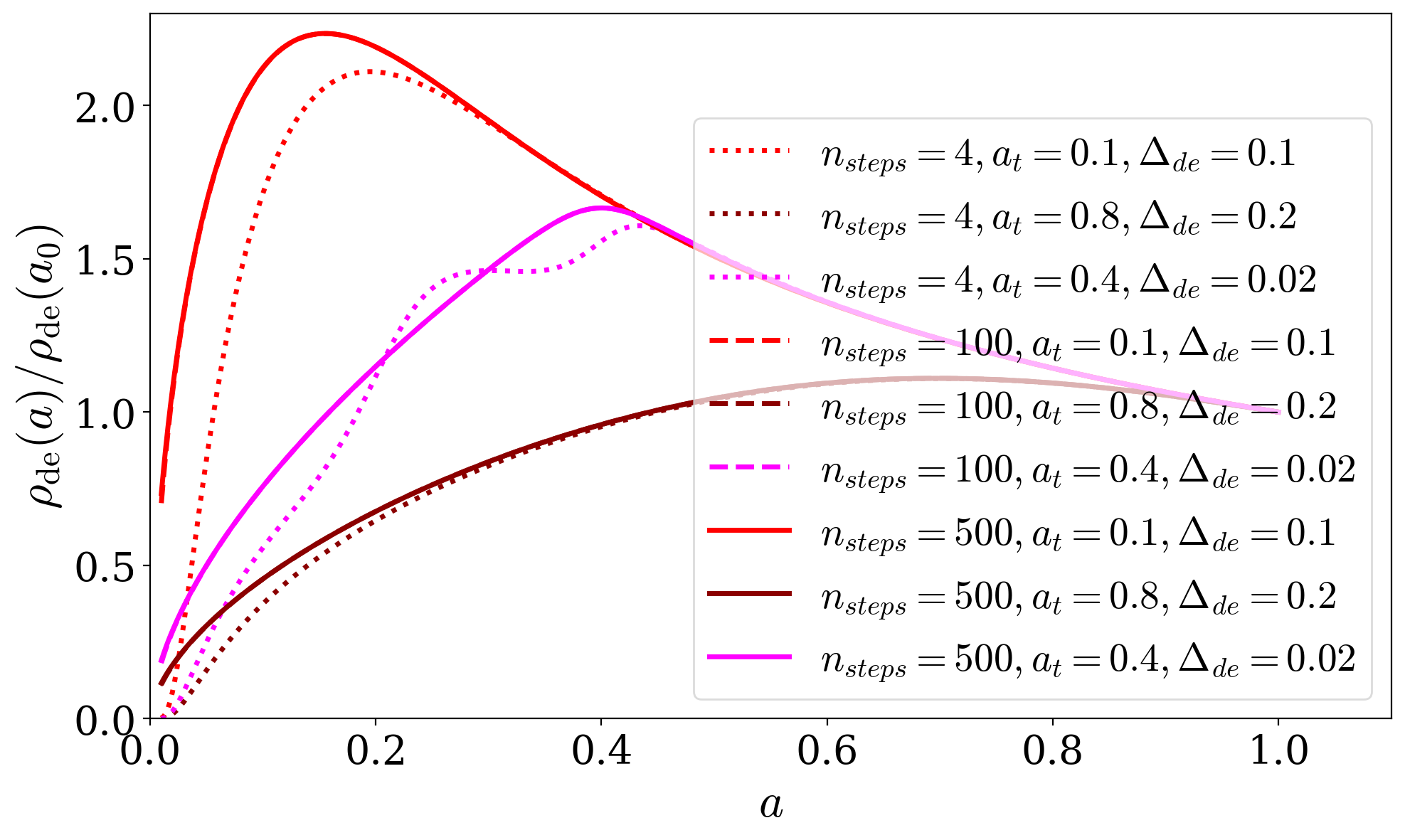}
    \caption{Evolution of $\frac{\rho_{\mathrm{de}}(a)}{\rho_{\mathrm{de}}(a_0)}$ for different sets of values for $n_{\mathrm{steps}}$, $a_{\mathrm{t}}$, and $\Delta_{\mathrm{de}}$, where the integral steps are set to $n_{\mathrm{steps}} = 4$, $100$, and $500$, respectively. The parameters $w_0$ and $w_{\mathrm{m}}$ are fixed to $-0.8$ and $-1.2$, respectively.  Note that the curves for $n_{\mathrm{steps}} = 100$ and $n_{\mathrm{steps}} = 500$ overlap, demonstrating that the integral converges
sufficiently at $n_{\mathrm{steps}} = 100$.}
    \label{fig:rho_steps}
\end{figure}

We consider the spatially flat Friedmann-Lema\^{i}tre-Robertson-Walker (FLRW) line element of our universe, which approximates it on large scales. This line element is given by $ds^2 = -dt^2 + a^2(t) (dx^2 + dy^2 + dz^2)$, in which $a(t)$ refers to the expansion scale factor of the universe in terms of the cosmic time $t$, and $(t, x, y, z)$ are the co-moving coordinates. Assuming GR in the background, the Hubble equation can be written as
$H^2 = (\kappa^2/3) \times (\rho_{\rm b} + \rho_{\rm r} + \rho_{\nu} + \rho_{\rm c} + \rho_{\rm de})$,
in which $\kappa^2$ is Einstein's gravitational constant
and $\rho_i$ is the energy density of the $i$-th component
(${\rm b}$ stands for baryons, ${\rm r}$ stands for radiation, ${\nu}$ stands for
neutrinos, ${\rm c}$ and ${\rm de}$ correspond to CDM and DE, respectively).
Under the assumption of no interaction between the fluids, one can determine the evolution of each individual component separately.\footnote{Let us note that as commonly adopted in the literature, the sum of the neutrino masses is fixed to be $0.06$~eV and the
number of neutrino species is fixed to $N_{\rm eff} = 3.044$.} In what follows, we focus on the evolution of the dark energy sector, whose equation of state is given by~\cite{Corasaniti:2002vg,Corasaniti:2004sz}:

\begin{eqnarray}\label{DE-EoS}
w_{\rm de} (a) = w_0 + (w_m - w_0) \mathcal{G} (a),
\end{eqnarray}
where $\mathcal{G} (a)$ is given by 
\begin{align}
    \mathcal{G} (a) = \frac{1 - \exp(- (a - 1)/\Delta_{\rm de})}{1 - \exp(1/\Delta_{\rm de})} \times \frac{1 + \exp(a_t/\Delta_{\rm de})}{1 + \exp(-(a - a_t)/\Delta_{\rm de})}.  
\end{align}
In the above description of the DE EoS, $w_0$ refers to the present-day value of the DE EoS, $w_m$ is the initial value of $w_{\rm de}$, i.e., $w_{\rm de} = w_m (a \ll 1)$, $a_t$ corresponds to the scale factor at which the transition from $w_m$ to $w_0$ occurs, and $\Delta_{\rm de}$ denotes the steepness of the transition. Therefore, eqn.~(\ref{DE-EoS}) contains four free parameters that need to be constrained. For the above EoS of DE, the energy density of the DE sector can be found from the following integral:

\begin{eqnarray}\label{rho-integral}
 \rho_{\rm de} = \rho_{\rm de,0}~ a^{-3} \times \exp\left( -3 \int_{a_0 = 1}^{a} \frac{w_{\rm de}(a')}{a'} ~da'\right).  
\end{eqnarray}
Now, considering the evolution of DE and other fluids present in the Hubble equation, one can, in principle, determine the expansion history of the universe at the background level. Figs.~\ref{fig:w_de} and~\ref{fig:rho_steps} provide a clear evolutionary history of the DE EoS, $w_{\rm de}(a)$, and its energy density in terms of $\rho_{\rm de}/\rho_{\rm de,0}$, considering different values of $a_t$ and $\Delta_{\rm de}$. We note that while computing the integral in (\ref{rho-integral}), we have used different integral steps $n_{\mathrm{steps}}$ in order to check the robustness achieved for a specific $n_{\mathrm{steps}}$.  
After comparing three different cases for $n_{\mathrm{steps}} = 4$,
$n_{\mathrm{steps}} = 100$, and $n_{\mathrm{steps}} = 500$,
we find that the evolution of $\rho_{\mathrm{de}}(a)$ for
$n_{\mathrm{steps}} = 100$ is indistinguishable from that obtained for
$n_{\mathrm{steps}} = 500$ (see Fig.~\ref{fig:rho_steps}).
This perfect overlap indicates that numerical convergence is reached; therefore,
the integral of $\rho_{\mathrm{de}}(a)$ is robust for
$n_{\mathrm{steps}} = 100$.\footnote{Considering the robustness, during the statistical analysis, we set $n_{\mathrm{steps}}=100$ inside CAMB~\cite{Lewis:1999bs}.}

We now turn our attention to the behavior of the model at the perturbative level.
Following~\cite{Ma:1995ey}, the evolution of all components at the perturbative
level can be derived straightforwardly.
Choosing the synchronous gauge, the perturbed version of the spatially flat
FLRW metric in terms of the conformal time $\tau$ reads~\cite{Ma:1995ey}
$ds^2 = a^2(\tau)\left[-d\tau^2 + (\delta_{ij} + h_{ij}) dx^i dx^j \right]$.
Taking into account the perturbed gravitational equations, one can then obtain
the evolution equations for the individual fluids appearing in the Hubble
equation (see also~\cite{Giare:2024gpk,Giare:2025pzu}).
In the above metric, $\delta_{ij}$ denotes the unperturbed spatial part of the
metric tensor, while $h_{ij}$ represents the perturbed spatial part.
In terms of the dimensionless density perturbation of the $i$-th fluid,
$\delta_i \equiv \delta \rho_i / \rho_i$, and the divergence of its velocity
field, $\theta_i \equiv i k^j v_j$, in Fourier space, the evolution equations
take the form
\begin{eqnarray}
\delta'_{i} & = & - (1+ w_{i})\left(\theta_{i}+ \frac{h'}{2}\right)
- 3\mathcal{H}\left(\frac{\delta P_i}{\delta \rho_i} - w_{i} \right)\delta_i
\nonumber \\ 
& & - 9 \mathcal{H}^2\left(\frac{\delta P_i}{\delta \rho_i} - c^2_{a,i} \right)
(1+w_i)\frac{\theta_i}{k^2}, \label{per1} \\
\theta'_{i} & = & - \mathcal{H}\left(1- 3 \frac{\delta P_i}{\delta \rho_i}\right)
\theta_{i}
+ \frac{\delta P_i/\delta \rho_i}{1+w_{i}}\, k^2\, \delta_{i}
- k^2\sigma_i, \label{per2}
\end{eqnarray}
where the prime denotes differentiation with respect to conformal time,
$\prime \equiv d/d\tau$, $\mathcal{H}(a)$ is the conformal Hubble parameter, and
$h$ is the usual synchronous-gauge metric perturbation.
Here, $k$ denotes the wavenumber in Fourier space.
The quantity $\sigma_i$ represents the anisotropic stress of the $i$-th fluid;
in this work, we set $\sigma_i = 0$.
Furthermore, $\delta P_i / \delta \rho_i$ defines the square of the sound speed
of the $i$-th fluid in its rest frame.
In particular, $\delta P_{\rm de} / \delta \rho_{\rm de} \equiv c^2_{\rm s,de}$
corresponds to the sound speed of the dark energy component.
The adiabatic sound speed is given by
$c^2_{a,i} = w_i - w_i^{\prime}/\!\left[3 \mathcal{H} (1+w_i)\right]$.
Following the existing literature, we set $c^2_{\rm s,de} = 1$, as appropriate
for minimally coupled scalar field models, and $c^2_{\rm s,c} = 0$.

\begin{figure*}[t!]
    \centering
    \begin{subfigure}[t]{0.49\textwidth}
        \centering
        \includegraphics[width=0.99\linewidth]{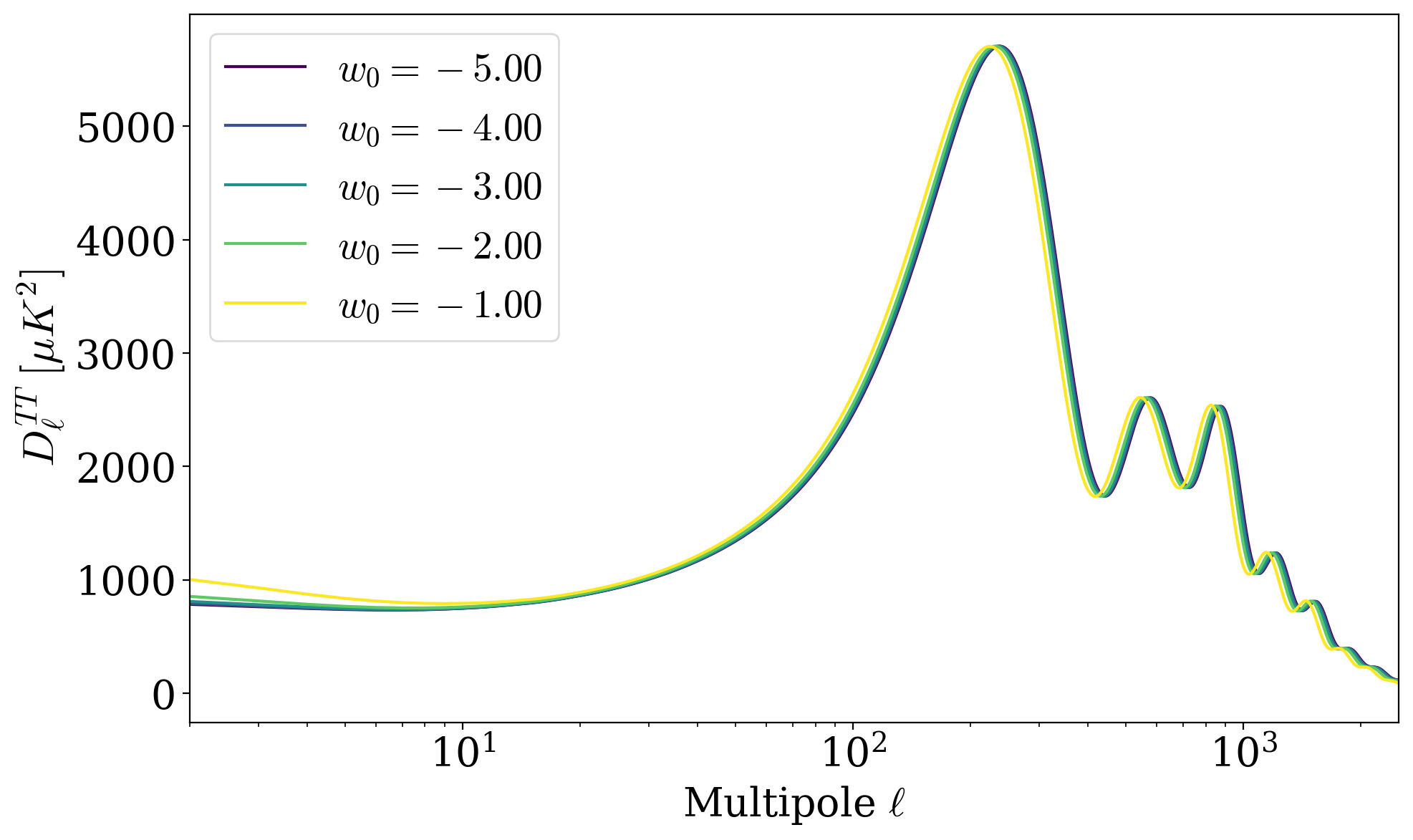}
    \end{subfigure}
    \begin{subfigure}[t]{0.49\textwidth}
        \centering
        \includegraphics[width=0.99\linewidth]{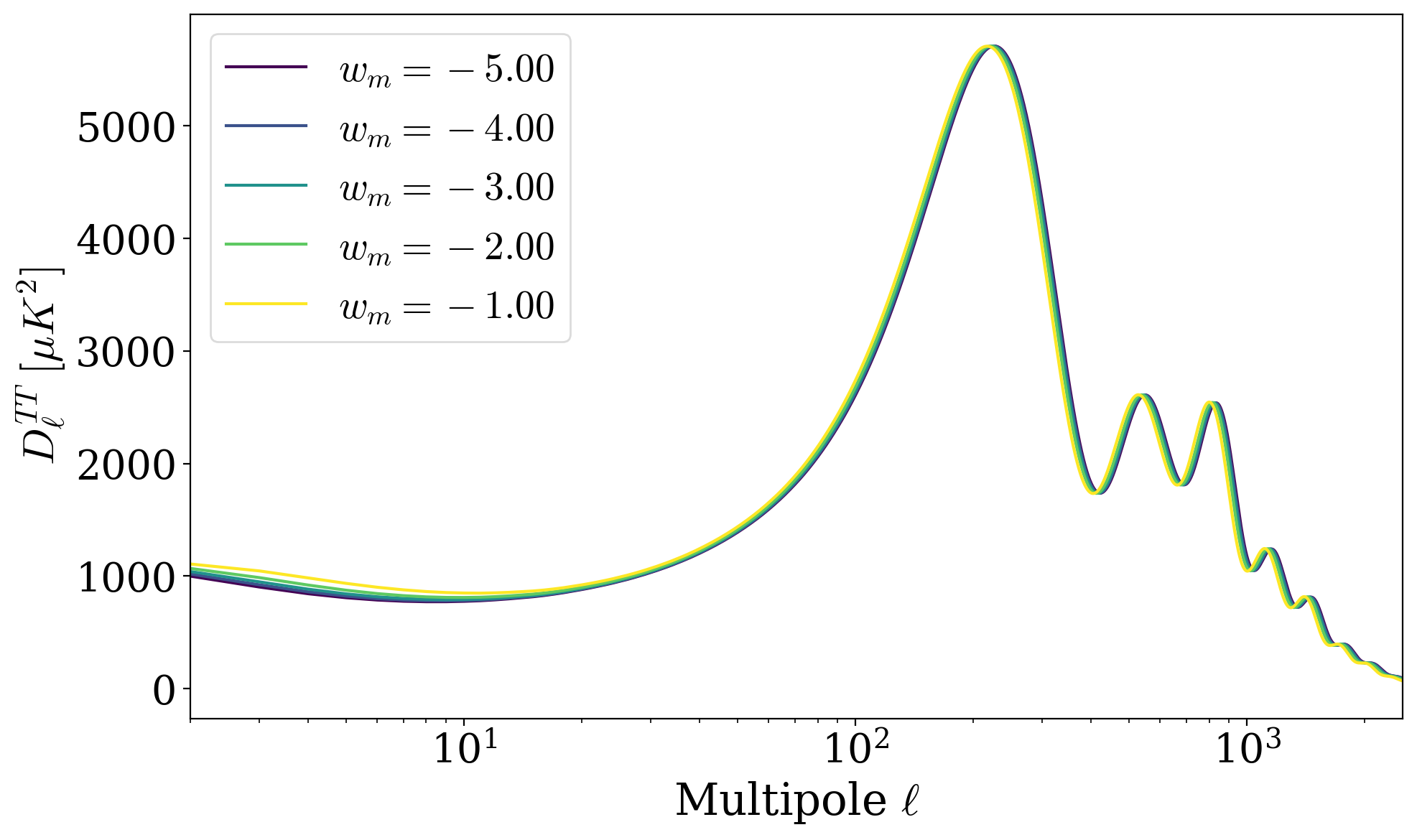}
    \end{subfigure}
    \begin{subfigure}[t]{0.49\textwidth}
        \centering
        \includegraphics[width=0.99\linewidth]{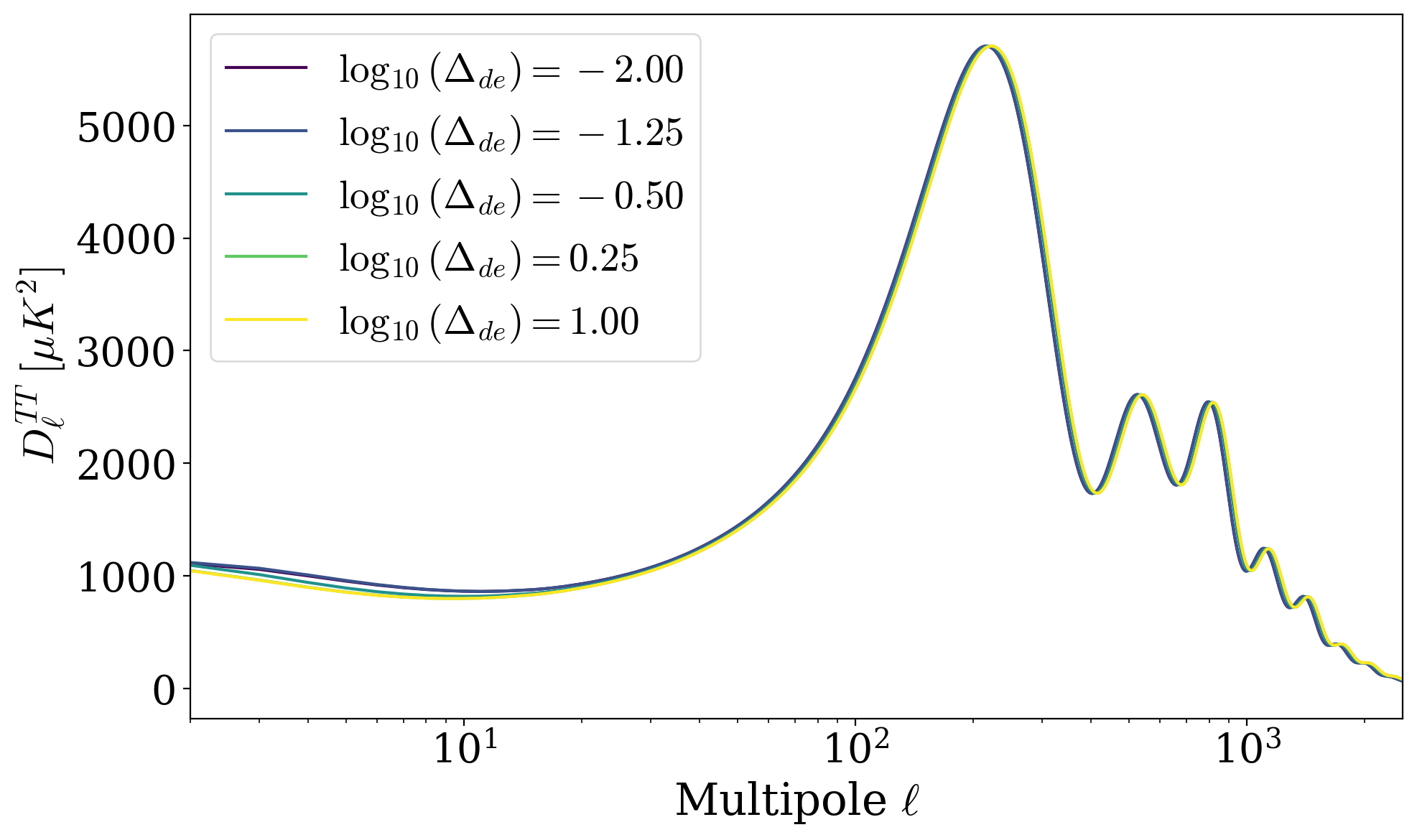}
    \end{subfigure}
    \begin{subfigure}[t]{0.49\textwidth}
        \centering
        \includegraphics[width=0.99\linewidth]{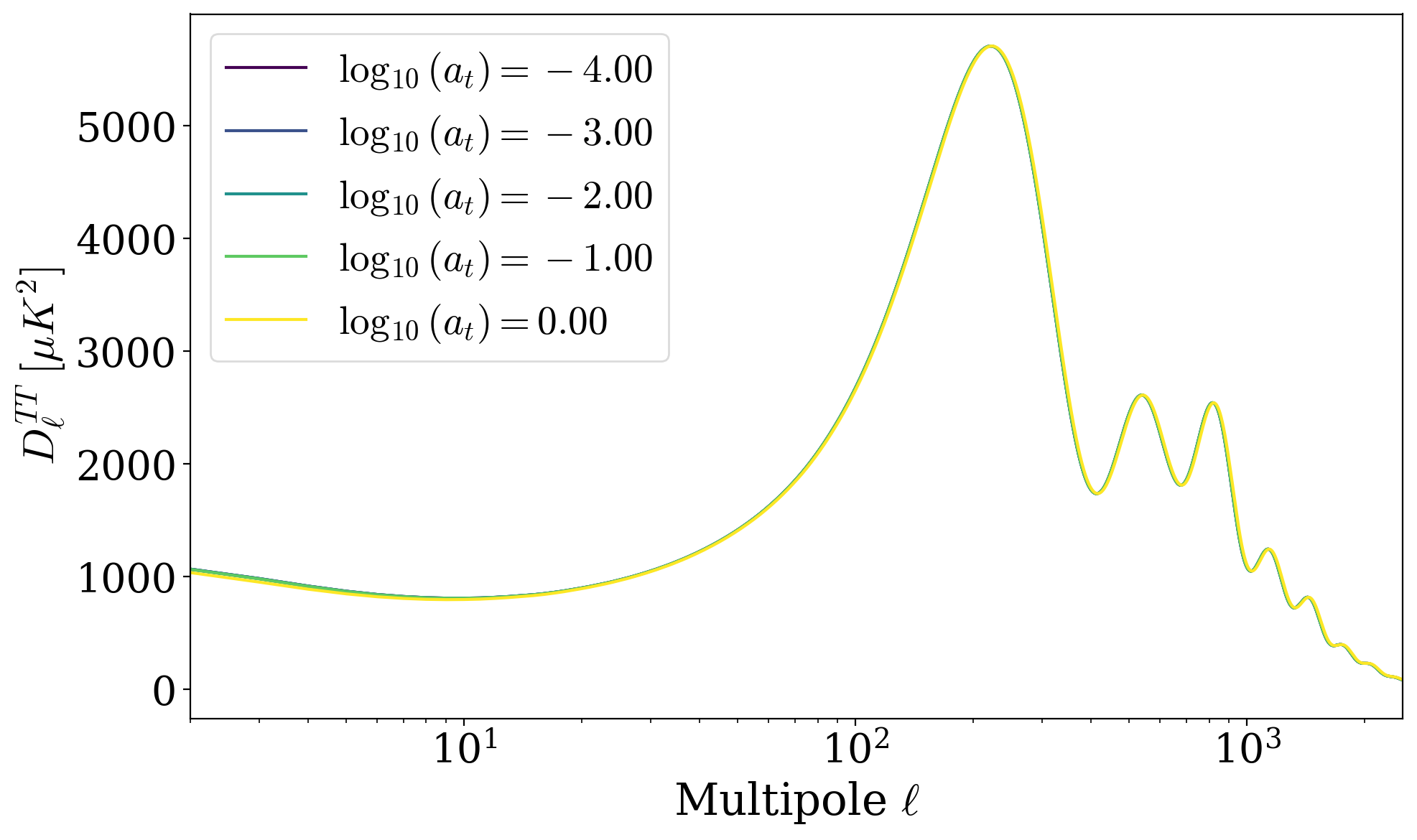}
    \end{subfigure}
    \caption{Theoretical prediction demonstrating the impact on the CMB TT power spectrum when varying different dark energy equation-of-state parameters (see the legends), while keeping the other EoS parameters fixed to the values $w_0 = -0.8$, $w_{\rm m} = -2.2$, $\log_{10}(\Delta_{\mathrm{de}}) = -0.2$, and $\log_{10}(a_\mathrm{t}) = -1$. Other cosmological parameters are fixed to the Planck 2018 best-fit values.} 
    \label{fig:CTT}
\end{figure*}

Finally, in Fig.~\ref{fig:CTT}, we show the impact on the CMB TT power spectrum when a particular DE parameter is varied while other parameters are fixed. We have shown four different plots where the four parameters of the DE EoS are individually varied. While drawing the plots, the other cosmological parameters are fixed to the Planck 2018 best-fit values using Planck 2018 TT,TE,EE+lowE+lensing~\cite{Planck:2018vyg}.
In the upper left plot of Fig.~\ref{fig:CTT}, we show the effects on the CMB TT spectrum for varying $w_0$ while fixing the other free parameters ($w_{\rm m} = -2.2$, $\log_{10}(\Delta_{\mathrm{de}}) = -0.2$, $\log_{10}(a_\mathrm{t}) = -1$). From this plot, one can see that changes in the low multipole region appear for different values of $w_0$. In particular, going deeply phantom leads to a suppression of the late-time ISW plateau. 
The upper right panel of Fig.~\ref{fig:CTT} shows the impact on the CMB TT power spectrum for different values of $w_m$, keeping the remaining free parameters fixed ($w_0 = -0.8$, $\log_{10}(\Delta_{\mathrm{de}}) = -0.2$, $\log_{10}(a_\mathrm{t}) = -1$). From this plot, one can see that there is no suppression of the ISW plateau, but we do observe a shift of the high-multipole peaks toward larger $\ell$ as $w_m$ becomes very negative, although less pronounced than in the $w_0$ case.
A similar effect is found in the lower left plot of Fig.~\ref{fig:CTT}, where we varied $\Delta_{\rm de}$ in terms of its logarithmic value, keeping the other parameters fixed ($w_0 = -0.8$, $w_{\rm m} = -2.2$, $\log_{10}(a_\mathrm{t}) = -1$). As $\log_{10}(\Delta_{\rm de})$ increases from negative to positive values, the same shift in the high-$\ell$ peaks is observed. This suggests a negative correlation between $\Delta_{\rm de}$ and $w_m$ in their effect on the spectrum.
In the lower right panel of Fig.~\ref{fig:CTT}, we illustrate the impact of the
transition scale factor $a_t$ on the CMB TT power spectrum by varying
$\log_{10}(a_t)$ while keeping the other dark energy parameters fixed
($w_0 = -0.8$, $w_{\rm m} = -2.2$, $\log_{10}(\Delta_{\mathrm{de}}) = -0.2$).
We find that variations in $a_t$ do not produce any noticeable changes in the
CMB TT spectrum, neither at low multipoles nor at high multipoles, even when the
magnitude of $\log_{10}(a_t)$ is significantly increased. This behavior reflects the fact that $a_t$ primarily controls the timing of the
dark energy transition, which occurs at late times when dark energy affects the
CMB only through the late Integrated Sachs--Wolfe effect.
For the range of values explored, the resulting evolution of the gravitational
potentials remains nearly unchanged, leading to a negligible impact on the CMB
anisotropies.
Consequently, $a_t$ remains largely unconstrained by current CMB data. 

\section{Observational data and statistical methodology}
\label{sec-data}

\begin{table}[h!]
\centering
\begin{tabular}{ccc}
\hline
Model & Parameter & Prior \\
\hline\hline
$\Lambda$CDM & $\Omega_{\mathrm{b}} h^2$ & $[0.005 , 0.1]$ \\
$\Lambda$CDM & $\Omega_{\mathrm{c}} h^2$ & $[0.001 , 0.99]$ \\
$\Lambda$CDM & $\tau$ & $[0.01 , 0.8]$ \\
$\Lambda$CDM & $100\,\theta_{\mathrm{s}}$ & $[0.5 , 10]$ \\
$\Lambda$CDM & $\ln(10^{10} A_{\mathrm{s}})$ & $[1.61 , 3.91]$ \\
$\Lambda$CDM & $n_{\mathrm{s}}$ & $[0.8 , 1.2]$ \\
\hline
4PDE & $w_0$ & $[-5.0 , 0.0]$ \\
4PDE & $w_{\mathrm{m}}$ & $[-5.0 , -0.5]$ \\
4PDE & $\log_{10}(\Delta_{\mathrm{de}})$  & $[-3.0 , 1.0]$ \\
4PDE & $\log_{10}(a_{\rm t})$  & $[-4.0 , 0.0]$ \\
\hline\hline
\end{tabular}
\caption{Flat prior distributions imposed on the cosmological parameters used in our analysis. The 4PDE models include the standard $\Lambda$CDM parameters along with four additional parameters.}
\label{tab:prior_table}
\end{table}

To perform parameter inference, we utilize the \texttt{Cobaya} tool~\cite{Torrado:2020dgo}, which implements a Markov Chain Monte Carlo (MCMC) sampler specifically designed for cosmological analyses. This is coupled with an adapted version of the \texttt{CAMB} Boltzmann solver~\cite{Lewis:1999bs}, modified to incorporate our Dynamic Dark Energy (DDE) parameterization. Perturbations in DE are modeled using the standard parameterized post-Friedmann (PPF) approach provided in \texttt{CAMB}~\cite{Lewis:1999bs}. The convergence of the MCMC chains is assessed by evaluating the Gelman–Rubin statistic $R - 1$~\cite{gelman_inference_1992}, with convergence accepted at $R - 1 < 0.02$. Posterior distributions and parameter contours are analyzed and visualized via the \texttt{getdist} package~\cite{Lewis:2019xzd}.

For a comprehensive comparison of the four-parameter DE models, we use the following observational datasets:
\begin{itemize}
\item \textbf{CMB:} Cosmic Microwave Background (CMB) measurements from the \textit{Planck} 2018 legacy data release, incorporating high-$\ell$ Plik TT, TE, and EE likelihoods, the low-$\ell$ TT-only Commander likelihood, and the low-$\ell$ EE-only SimAll likelihood~\cite{Planck:2018nkj,Planck:2019nip,Planck:2018lbu}, combined with the Planck 2018 lensing likelihood~\cite{Planck:2018lbu}. This integrated dataset is collectively referred to as \textbf{CMB}.

\item \textbf{BAO:} Baryon Acoustic Oscillation (BAO) measurements from the initial three years of the Dark Energy Spectroscopic Instrument (DESI DR2)~\cite{DESI:2025zgx,DESI:2025fii,DESI:2025qqy}, labelled as \textbf{DESI}.

\item \textbf{Type Ia Supernovae:} Distance modulus observations from Type Ia Supernovae (SNeIa) sourced from the PantheonPlus compilation~\cite{Scolnic:2021amr,Brout:2022vxf}, encompassing 1701 light curves from 1550 distinct SNeIa spanning the redshift range $z \in [0.001,\, 2.26]$, referred to as \textbf{PantheonPlus}. Additionally, we incorporate the full five-year dataset from the Dark Energy Survey (DES), comprising 1635 SNeIa across redshifts $0.1 < z < 1.13$~\cite{DES:2024hip,DES:2024jxu,DES:2024upw}, referred to as \textbf{DESY5}, and the Union3 compilation, comprising 2087 SNe~\cite{Rubin:2023ovl}, denoted as \textbf{Union3}.
\end{itemize}

We apply uniform flat priors as specified in Table~\ref{tab:prior_table}. The expanded four-parameter DE models are structured as extensions of the standard six-parameter $\Lambda$CDM model. This baseline consists of the baryon density $\Omega_b h^2$, cold dark matter density $\Omega_c h^2$, reionization optical depth $\tau_{\mathrm{reio}}$, scalar amplitude and spectral index $\ln(10^{10} A_{\mathrm{s}})$ and $n_{\mathrm{s}}$, and the angular size at recombination of the sound horizon $\theta_s$. To this baseline, we append four additional dark energy equation-of-state parameters, detailed in Section~\ref{sec-2}. 

To quantify the statistical performance of the extended four-parameter DE (4PDE) model relative to the standard $\Lambda$CDM model, we evaluate the differences in the minimum chi-square values. We note that the total $\chi^2$ for each model is obtained by summing the
contributions from the individual datasets (CMB, BAO, and SN).
In statistical parlance, combining these likelihoods under the assumption of
independence constitutes a composite likelihood approach~\cite{Varin:2011,Lindsay:2011}.
While this method formally approximates the full joint distribution, it is well
motivated in this context given the distinct physical regimes probed and the
negligible cross-correlations among the selected datasets.
The difference is defined as:
\begin{equation}
    \label{eq:delta-chi-square}
    \Delta \chi^2_{\mathrm{min},\Lambda \mathrm{CDM}/\mathrm{CPL}} = \chi^2_{\mathrm{min},\mathrm{4PDE}} - \chi^2_{\mathrm{min},\Lambda \mathrm{CDM}/\mathrm{CPL}}.
\end{equation}
A negative value of the difference in eq.~(\ref{eq:delta-chi-square}) indicates that the data favor the extended 4PDE model over the $\Lambda$CDM or CPL model. Furthermore, we perform Bayesian model comparison by computing the logarithm of the Bayesian evidence $\ln \mathcal{Z}$ using \texttt{MCEvidence}~\cite{Heavens:2017afc}, interfaced through the \texttt{Cobaya} wrapper available in the \texttt{wgcosmo} repository~\cite{giare2025wgcosmo}. According to Bayes’ theorem, for each model $\mathcal{M}_i$ characterized by parameters $\Theta$, the posterior distribution is:
\begin{equation}
P(\Theta|D, \mathcal{M}_i) = \frac{\mathcal{L}(D|\Theta, \mathcal{M}_i) \, \pi(\Theta|\mathcal{M}_i)}{\mathcal{Z}_i},
\label{eq:bayes_theorem}
\end{equation}
where $\mathcal{L}$ denotes the maximum likelihood, $\pi$ represents the prior, and the evidence $\mathcal{Z}_i$ is computed as:
\begin{equation}
\mathcal{Z}_i = \int \mathcal{L}(D|\Theta, \mathcal{M}_i) \, \pi(\Theta|\mathcal{M}_i) \, {\rm d}\Theta.
\label{eq:bayesian_evidence}
\end{equation}

For model comparison, we calculate the Bayes factor $\mathcal{Z}_{\Lambda \mathrm{CDM}/\mathrm{CPL}} = \mathcal{Z}_{\mathrm{4PDE}} / \mathcal{Z}_{\Lambda \mathrm{CDM}/\mathrm{CPL}}$, and define the relative log-evidence as:
\begin{eqnarray}
\Delta \ln \mathcal{Z}_{\Lambda \mathrm{CDM}/\mathrm{CPL}} \equiv \ln \mathcal{Z}_{\mathrm{4PDE}} - \ln \mathcal{Z}_{\Lambda \mathrm{CDM}/\mathrm{CPL}},
\label{eq:relative_log_bayesian_evidence}
\end{eqnarray}
Positive values of $\Delta \ln \mathcal{Z}_{\Lambda \mathrm{CDM}/\mathrm{CPL}}$ indicate support for the extended model.

We interpret $\Delta \ln \mathcal{Z}_{\Lambda \mathrm{CDM}/\mathrm{CPL}}$ using the updated Jeffreys’ scale~\cite{Kass:1995loi}: ranges of $[0, 1]$ imply \textit{inconclusive} evidence, $[1, 2.5]$ suggest \textit{weak} evidence, $[2.5, 5]$ denote \textit{moderate} evidence, $[5, 10]$ reflect \textit{strong} evidence, and values exceeding $10$ represent \textit{very strong} evidence supporting the favored model.

\begin{table*}[htbp]
\scriptsize
\centering
\resizebox{\textwidth}{!}{%
\begin{tabular}{l|ccccccc}
\toprule
\textbf{Parameters} 
& \textbf{CMB} 
& \textbf{CMB} 
& \textbf{CMB}
& \textbf{CMB}
& \textbf{CMB+DESI}
& \textbf{CMB+DESI}
& \textbf{CMB+DESI} \\
& \textbf{+DESY5} 
& \textbf{+PantheonPlus} 
& \textbf{+Union3}
& \textbf{+DESI}
& \textbf{+DESY5}
& \textbf{+PantheonPlus}
& \textbf{+Union3} \\
\midrule
$\Omega_{\mathrm{b}}\,h^2$ 
& $0.02241 \pm 0.00015$ 
& $0.02240 \pm 0.00015$ 
& $0.02241 \pm 0.00014$
& $0.02241 \pm 0.00014$ 
& $0.02243 \pm 0.00013$
& $0.02245 \pm 0.00014$
& $0.02244 \pm 0.00013$ \\[6pt]

$\Omega_{\mathrm{c}}\,h^2$ 
& $0.1196 \pm 0.0012$ 
& $0.1196 \pm 0.0012$ 
& $0.1195 \pm 0.0012$
& $0.11948 \pm 0.00095$ 
& $0.11915 \pm 0.00091$
& $0.11899 \pm 0.00095$
& $0.11920 \pm 0.00090$ \\[6pt]

$100\,\theta_{\mathrm{MC}}$ 
& $1.04096 \pm 0.00032$ 
& $1.04097 \pm 0.00031$ 
& $1.04097 \pm 0.00031$
& $1.04099 \pm 0.00029$ 
& $1.04101 \pm 0.00028$
& $1.04106 \pm 0.00028$
& $1.04101 \pm 0.00028$ \\[6pt]

$\tau_{\mathrm{reio}}$ 
& $0.0526 \pm 0.0076$ 
& $0.0533^{+0.0067}_{-0.0076}$ 
& $0.0524 \pm 0.0072$
& $0.0522 \pm 0.0073$ 
& $0.0537 \pm 0.0071$
& $0.0547 \pm 0.0074$
& $0.0534 \pm 0.0071$ \\[6pt]

$n_{\mathrm{s}}$ 
& $0.9662 \pm 0.0041$ 
& $0.9663 \pm 0.0041$ 
& $0.9664 \pm 0.0041$
& $0.9667 \pm 0.0037$ 
& $0.9673 \pm 0.0037$
& $0.9678 \pm 0.0037$
& $0.9673 \pm 0.0037$ \\[6pt]

$\ln\bigl(10^{10}\,A_{\mathrm{s}}\bigr)$ 
& $3.040 \pm 0.015$ 
& $3.041 \pm 0.014$ 
& $3.039 \pm 0.014$
& $3.039 \pm 0.014$ 
& $3.041 \pm 0.014$
& $3.043 \pm 0.014$
& $3.041 \pm 0.014$ \\[6pt]

$w_0$ 
& $-0.820^{+0.099}_{-0.14}$ 
& $-0.927^{+0.032}_{-0.083}$ 
& $-0.74^{+0.15}_{-0.19}$
& $-0.57 \pm 0.24$ 
& $-0.804 \pm 0.066$
& $-0.899^{+0.050}_{-0.062}$
& $-0.74 \pm 0.10$ \\[6pt]

$w_{\mathrm{m}}$ 
& $-2.39^{+1.3}_{-0.65}$ 
& $< -1.15$ 
& $-2.57^{+1.3}_{-0.73}$
& $-2.41^{+0.92}_{-0.52}$ 
& $-2.18^{+1.1}_{-0.44}$
& $-2.58^{+1.6}_{-0.89}$
& $-2.28^{+1.0}_{-0.35}$ \\[6pt]

$\log_{10}(\Delta_{\mathrm{de}})$ 
& $---$ 
& $---$ 
& $---$
& $> -1.03$ 
& $> -0.896$
& $-0.63 \pm 0.70$
& $-0.15^{+0.49}_{-0.71}$ \\[6pt]

$\log_{10}(a_{\rm t})$ 
& $---$ 
& $---$ 
& $---$
& $---$ 
& $---$
& $---$
& $---$ \\[6pt]
\midrule

$H_0$ [km/s/Mpc] 
& $67.0 \pm 1.2$ 
& $67.50^{+0.94}_{-1.5}$ 
& $66.8 \pm 1.5$
& $64.8^{+2.1}_{-2.4}$ 
& $66.94 \pm 0.56$
& $67.71 \pm 0.59$
& $66.27 \pm 0.87$ \\[6pt]

$\sigma_{8}$ 
& $0.807 \pm 0.013$ 
& $0.811^{+0.011}_{-0.014}$ 
& $0.805 \pm 0.015$
& $0.788^{+0.018}_{-0.020}$ 
& $0.8048 \pm 0.0087$
& $0.8105 \pm 0.0091$
& $0.799 \pm 0.010$ \\[6pt]

$S_{8}$ 
& $0.831 \pm 0.012$ 
& $0.829 \pm 0.012$ 
& $0.832 \pm 0.012$
& $0.839 \pm 0.014$ 
& $0.8279 \pm 0.0094$
& $0.8238 \pm 0.0098$
& $0.831 \pm 0.010$ \\[6pt]

$\Omega_{\mathrm{m}}$ 
& $0.319 \pm 0.013$ 
& $0.314^{+0.014}_{-0.011}$ 
& $0.320 \pm 0.015$
& $0.341 \pm 0.024$ 
& $0.3175 \pm 0.0056$
& $0.3099 \pm 0.0057$
& $0.3242 \pm 0.0088$ \\[6pt]
\midrule

$\Delta \chi^2_{min,\Lambda{\rm CDM}}$ 
& $-6.95$ 
& $-2.62$ 
& $-4.83$
& $-9.96$ 
& $-18.54$
& $-10.91$
& $-13.58$ \\[6pt]

$\Delta \ln \mathcal{Z}_{\Lambda{\rm CDM}}$ 
& $-1.92$ 
& $-4.75$ 
& $-2.52$
& $-1.25$ 
& $2.57$
& $-1.45$
& $0.67$ \\[6pt]

$\Delta \chi^2_{min,{\rm CPL}}$
& $0.46$
& $-0.33$
& $0.23$
& $0.47$
& $0.19$
& $-3.74$
& $0.74$ \\[6pt]

$\Delta \ln \mathcal{Z}_{{\rm CPL}}$
& $-0.54$
& $0.27$
& $-0.65$
& $-0.38$
& $0.03$
& $1.14$
& $-0.24$ \\[6pt]

\bottomrule
\end{tabular}
}
\caption{Observational constraints at 68\% CL and upper or lower limits at 95\% CL using various datasets. Here, $\Delta \chi^2_{\mathrm{min}}$ and $\Delta \ln \mathcal{Z}$ are defined as
$\Delta \chi^2_{\mathrm{min},\Lambda \mathrm{CDM}/\mathrm{CPL}}
= \chi^2_{\mathrm{min},\mathrm{4PDE}}
- \chi^2_{\mathrm{min},\Lambda \mathrm{CDM}/\mathrm{CPL}}$
and
$\Delta \ln \mathcal{Z}_{\Lambda \mathrm{CDM}/\mathrm{CPL}}
\equiv \ln \mathcal{Z}_{\mathrm{4PDE}}
- \ln \mathcal{Z}_{\Lambda \mathrm{CDM}/\mathrm{CPL}}$.
Negative values of
$\Delta \chi^2_{\mathrm{min},\Lambda \mathrm{CDM}/\mathrm{CPL}}$
favor the 4PDE model over the standard $\Lambda$CDM/CPL scenario, while positive
values of $\Delta \ln \mathcal{Z}_{\Lambda \mathrm{CDM}/\mathrm{CPL}}$ indicate a
preference for the 4PDE model.} 
\label{table-4-parameter-DDE}
\end{table*}

\section{Results and their implications}
\label{sec-results}

\begin{figure}[htbp]
    \centering
    \includegraphics[width=1.0\linewidth]{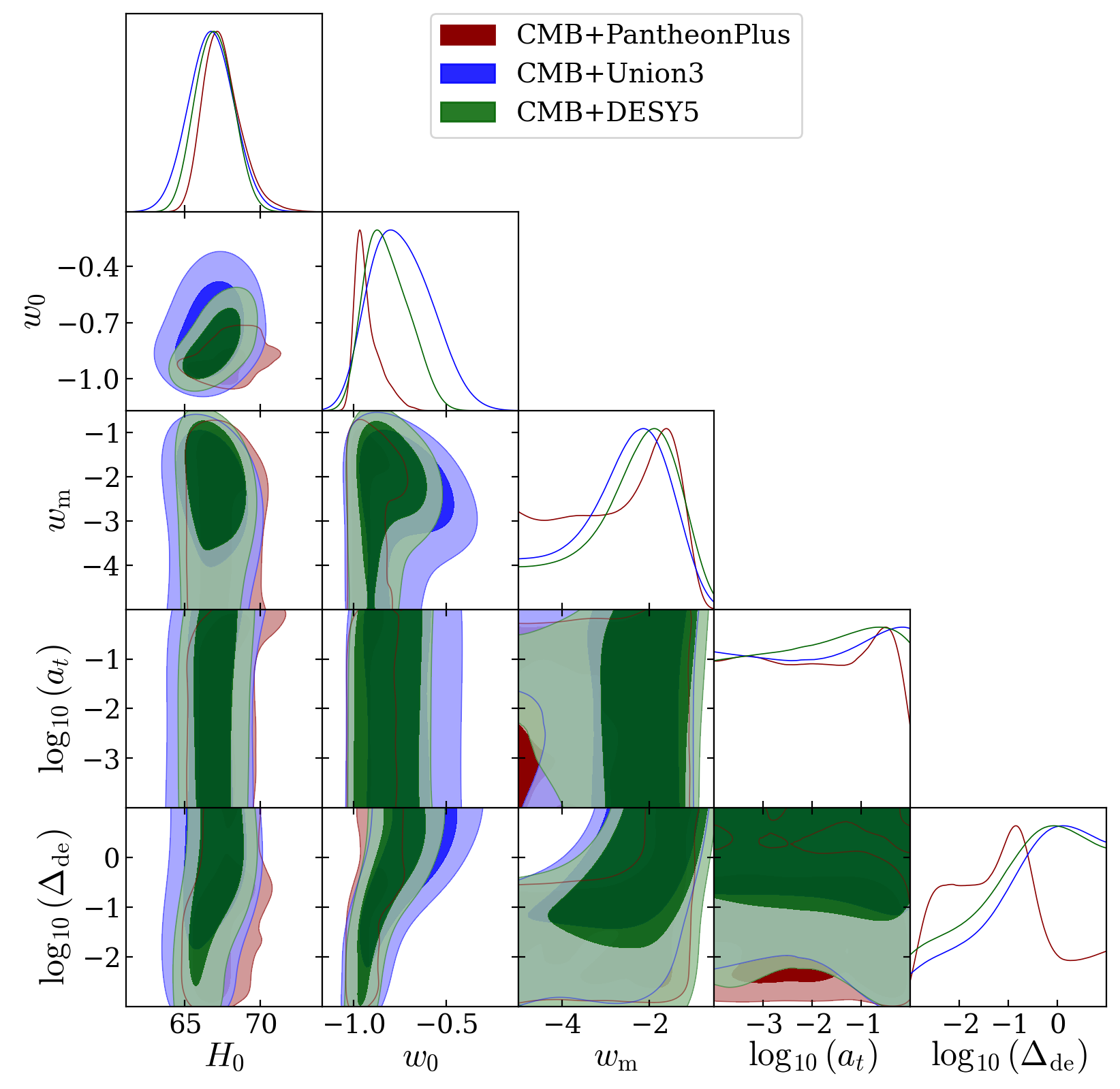}
    \caption{One-dimensional posterior distributions and two-dimensional marginalized contours for $H_0$ and the four model parameters, as obtained from the CMB+PantheonPlus, CMB+Union3, and CMB+DESY5 dataset combinations.}
    \label{fig:post_sn}
\end{figure}

\begin{figure}[htbp]
    \centering
    \includegraphics[width=1.0\linewidth]{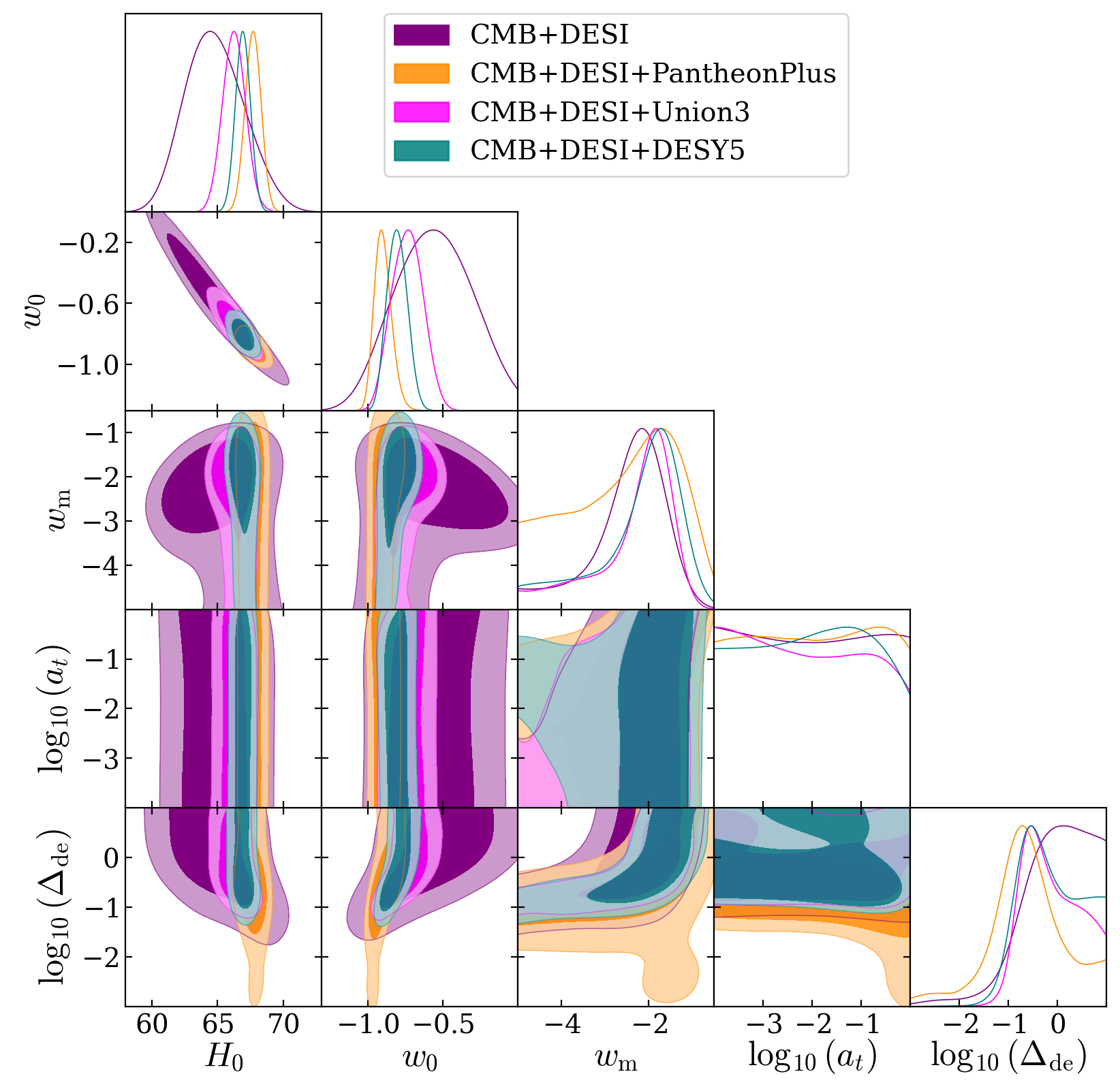}
    \caption{One-dimensional posterior distributions and two-dimensional marginalized contours for $H_0$ and the four model parameters, as obtained from the CMB+DESI, CMB+DESI+PantheonPlus, CMB+DESI+Union3, and CMB+DESI+DESY5 dataset combinations.}
    \label{fig:post_desi_sn}
\end{figure}

\begin{figure}[htbp]
    \centering
    \includegraphics[width=0.99\linewidth]{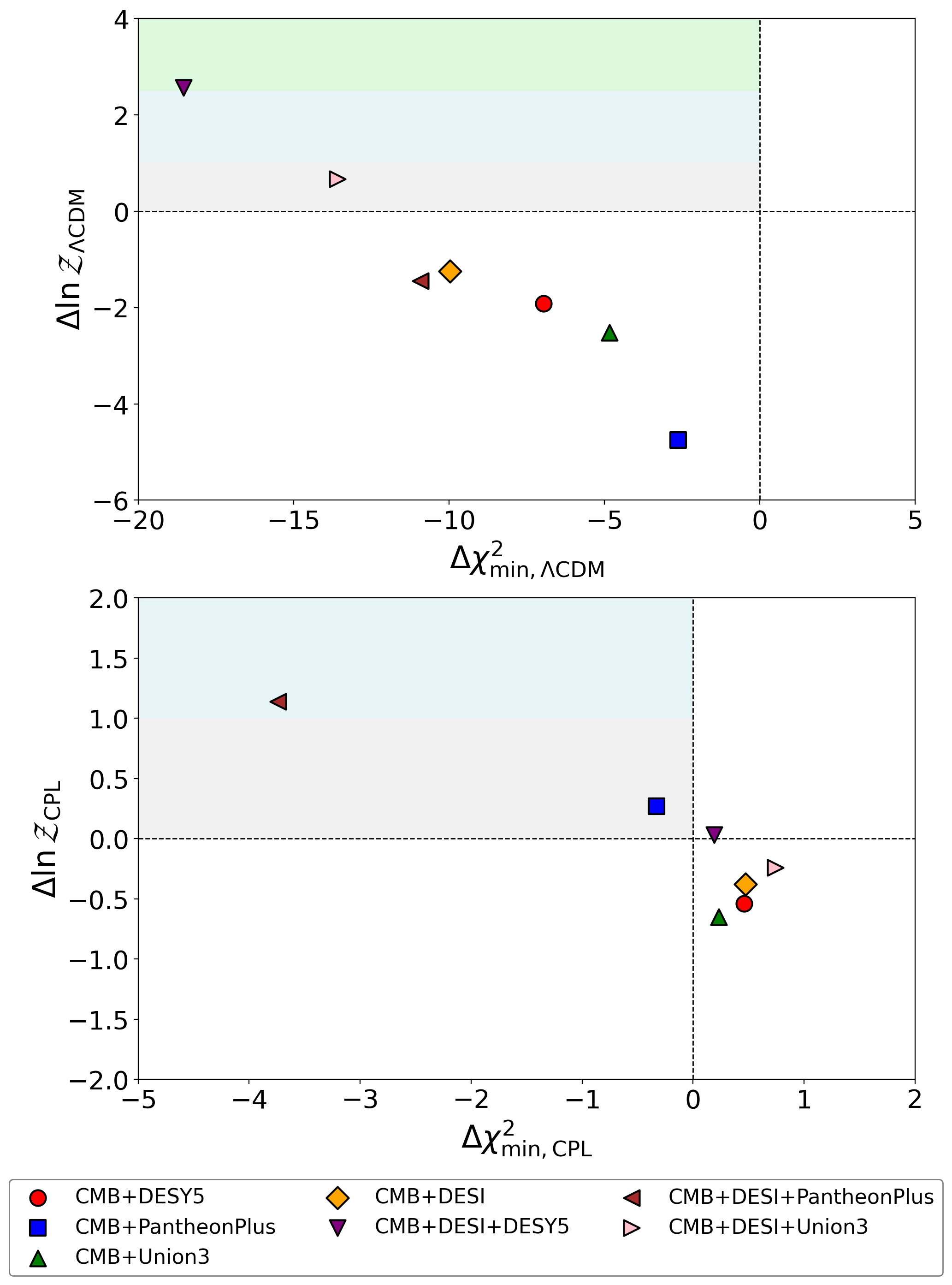}
    \caption{$\Delta \chi^2_{\mathrm{min},\Lambda{\rm CDM}}$ \textit{vs.} $\Delta \ln \mathcal{Z}_{\Lambda{\rm CDM}}$  and $\Delta \chi^2_{\mathrm{min},\rm CPL}$ \textit{vs.} $\Delta \ln \mathcal{Z}_{\rm CPL}$  for the 4PDE model, considering all datasets.}
    \label{fig:lnZ_chi2}
\end{figure}

\begin{figure*}[t!]
    \centering
    \begin{subfigure}[t]{0.49\textwidth}
        \centering
        \includegraphics[width=0.98\linewidth]{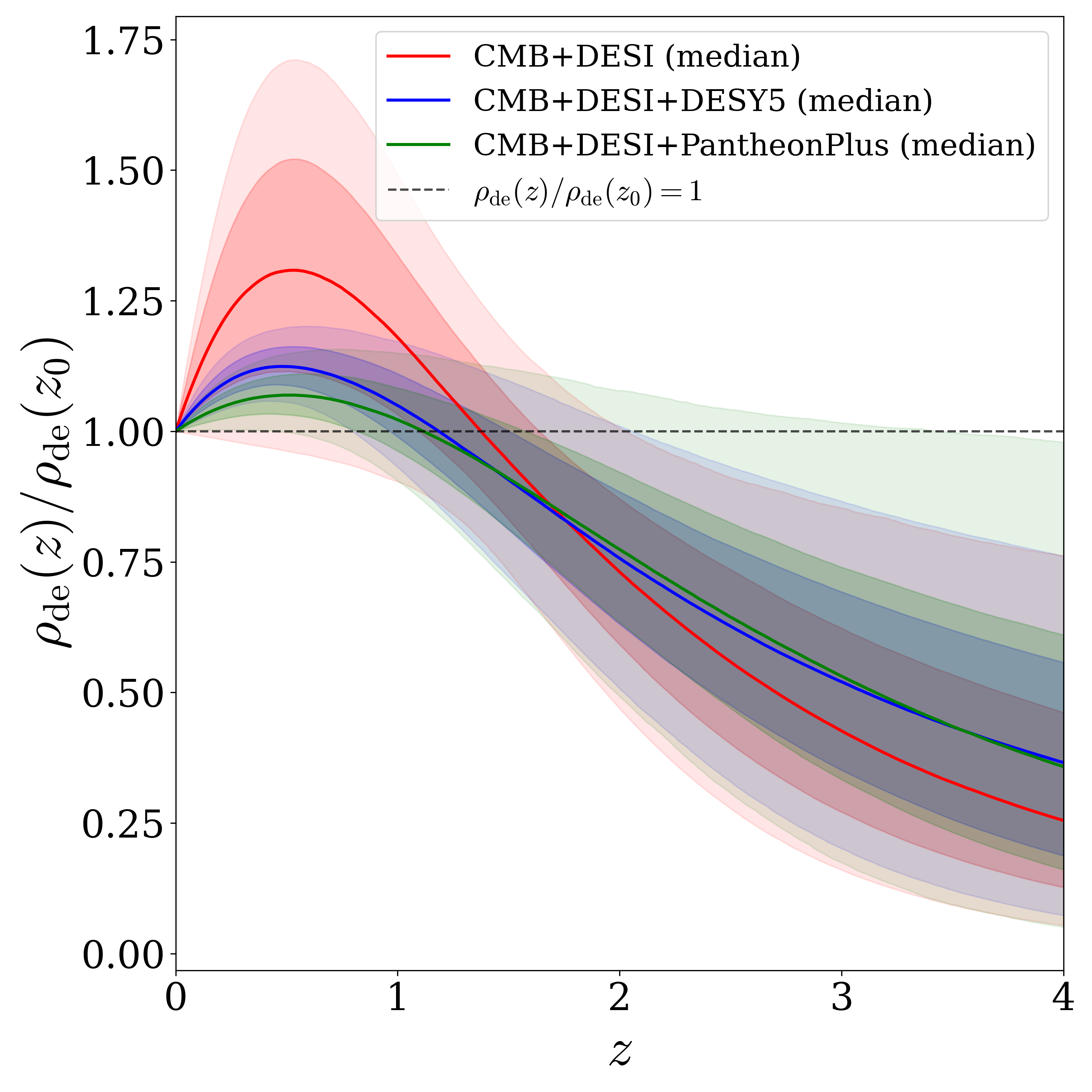}
    \end{subfigure}
    \begin{subfigure}[t]{0.49\textwidth}
        \centering
\includegraphics[width=0.98\linewidth]{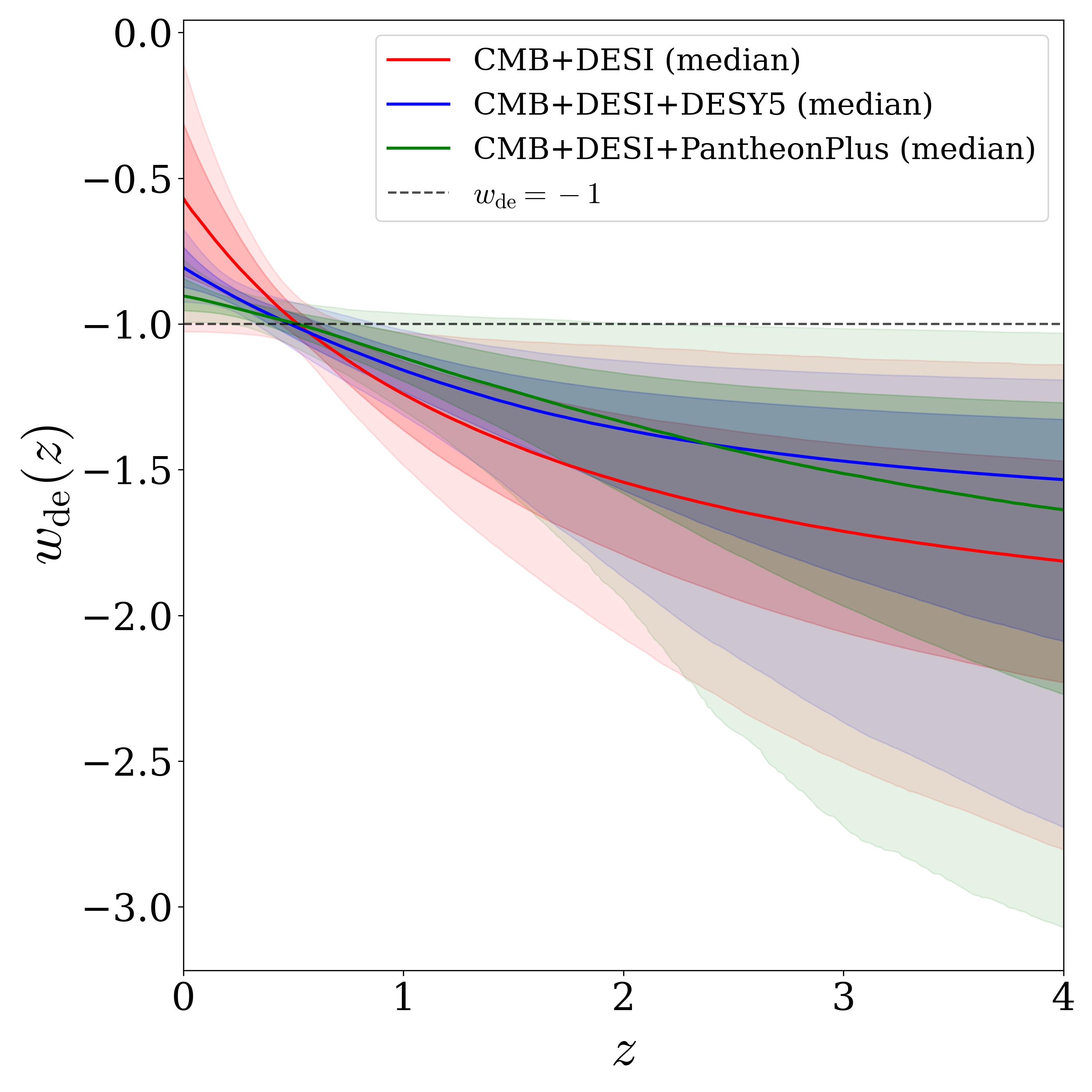}
    \end{subfigure}
    \caption{
    Evolution of the normalized DE density $\rho_{\mathrm{de}}(z)/\rho_{\mathrm{de}}(z_0)$ (left panel) and the DE EoS $w_{\rm de}(z)$ (right panel) as functions of the redshift $z$, based on MCMC samples using the combined datasets CMB+DESI, CMB+DESI+DESY5, and CMB+DESI+PantheonPlus, where $\rho_{\mathrm{de}}(z_0)$ is the DE density at the present day ($z=0$).}
    \label{fig:Reconstruction}
\end{figure*}

This section is devoted to the observational constraints of the 4PDE model. To this end, we use the latest cosmological probes, namely CMB, BAO, and three different compilations of SNeIa (DESY5, PantheonPlus, and Union3), and perform various combined analyses. Additionally, we compare the present model with respect to the $\Lambda$CDM model in terms of $\Delta \chi^2_{\mathrm{min},\Lambda{\rm CDM}}$ and $\Delta \ln \mathcal{Z}_{\Lambda{\rm CDM}}$. In Table~\ref{table-4-parameter-DDE}, we present the constraints on the free and derived parameters of this model, and in Figs.~\ref{fig:post_sn},~\ref{fig:post_desi_sn},~\ref{fig:Reconstruction}, and~\ref{fig:lnZ_chi2}, we graphically illustrate its behavior. As the number of free parameters in this model is significantly larger than in common DE parameterizations, the constraining power of CMB data alone is limited. Therefore, instead of performing a CMB-only analysis, we initially consider two combinations: CMB+SNeIa and CMB+DESI. We then perform a full analysis using the combined dataset CMB+DESI+SNeIa to improve parameter constraints. In the following, we discuss the results obtained from each combined dataset.

\begin{figure}[htbp]
    \centering
    \includegraphics[width=0.99\linewidth]{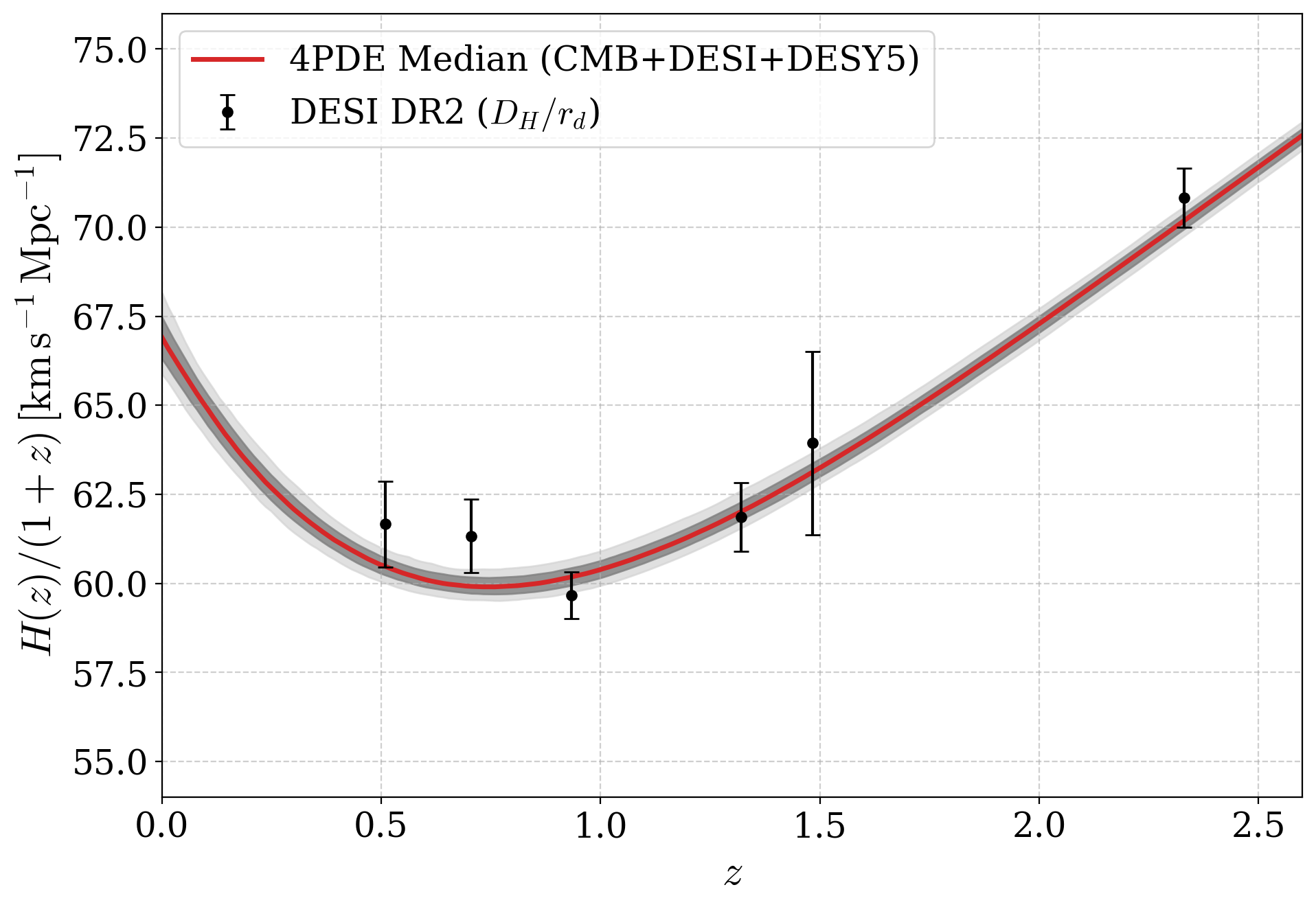}
    \caption{Comoving Hubble parameter, shown as $H(z)/(1+z)$, as a function of redshift $z$. The black data points with error bars are from DESI DR2, derived from the ratio of the Hubble distance to the sound horizon, $D_H/r_d$. The solid red line indicates the median curve, and the grey bands show the 68\% and 95\% confidence ranges allowed by CMB+DESI+DESY5 in the 4PDE model.} 
    \label{fig:comoving_H}
\end{figure}
\begin{figure}[htbp]
    \centering
    \includegraphics[width=0.99\linewidth]{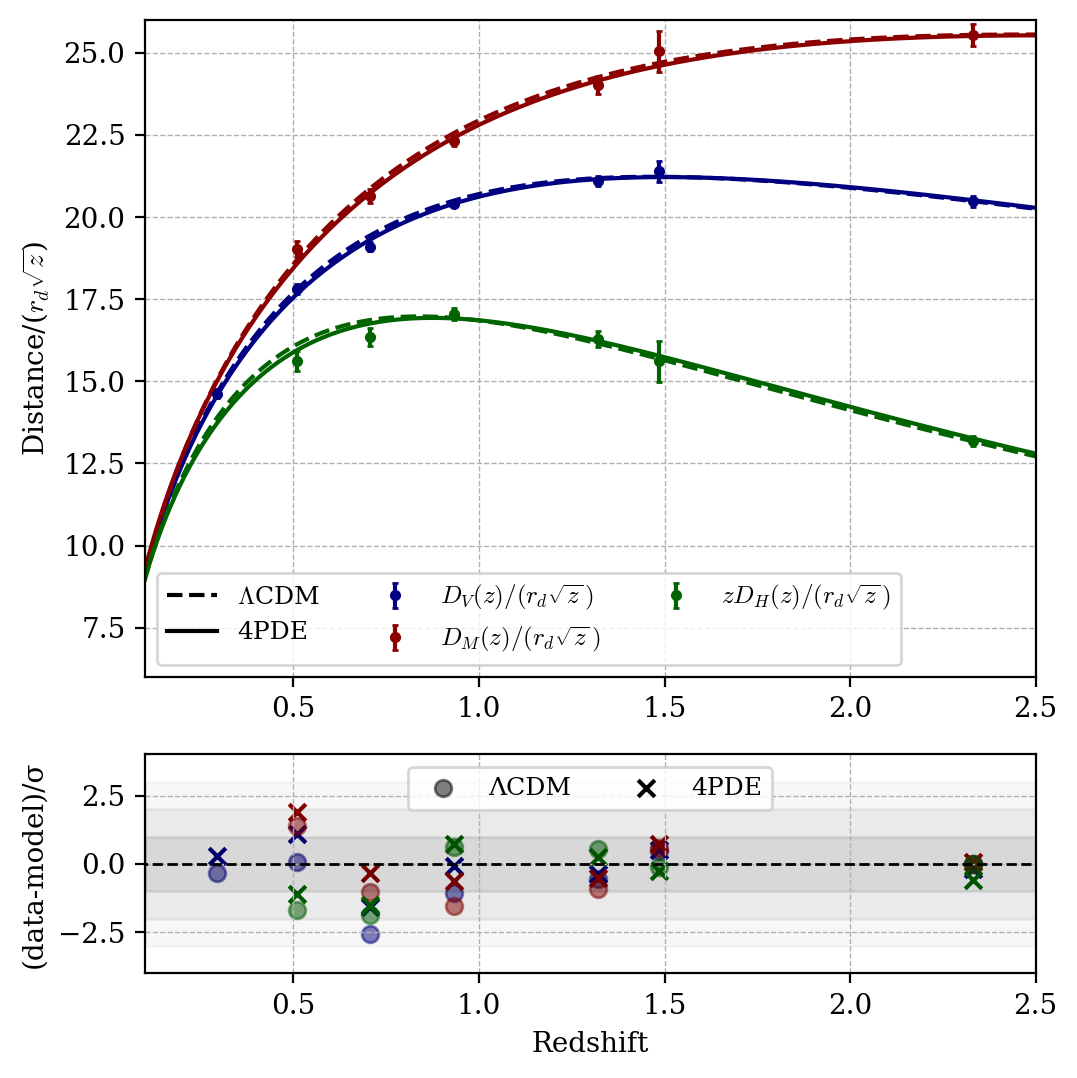}
    \caption{\textbf{Top panel:} BAO distance measurements. The solid lines represent the best-fit 4PDE model, while the dashed lines depict the best-fit standard $\Lambda$CDM model. Both models are constrained using the CMB+DESI+DESY5 dataset combination. \textbf{Bottom panel:} The residuals for each model, showing the difference between the data and the model predictions, normalized by the measurement uncertainty $\sigma$. Circles correspond to the $\Lambda$CDM model, and crosses correspond to the 4PDE model.}
    \label{fig:bao_D}
\end{figure}

We begin with the first three analyses: CMB+DESY5, CMB+PantheonPlus, and CMB+Union3. Examining the results on the free and derived parameters, our first observation is that the parameter $w_0$, which represents the present-day value of the DE EoS, lies in the quintessence regime at slightly more than 68\% CL for both CMB+DESY5 and CMB+Union3. In the case of CMB+PantheonPlus, although the mean value of $w_0$ also falls in the quintessence regime, it remains consistent with a cosmological constant $w_0 = -1$ within the 68\% CL ($w_0 = -0.927^{+0.032}_{-0.083}$).
The parameter $w_m$, denoting the initial value of the DE EoS, shows a phantom behavior at more than 68\% CL in the CMB+DESY5 and CMB+Union3 combinations ($w_m = -2.39^{+1.3}_{-0.65}$ and $w_m = -2.57^{+1.3}_{-0.73}$, respectively). For CMB+PantheonPlus, however, $w_m$ is only weakly constrained, with an upper limit $w_m < -1.15$ at 95\% CL. The possibility of phantom DE at early times is also supported by recent DESI results~\cite{DESI:2024aqx,DESI:2025fii}.
On the other hand, the parameters $a_t$ (the scale factor at which the transition from $w_m$ to $w_0$ occurs) and $\Delta_{\rm de}$ (the steepness of the transition) remain unconstrained by these datasets, as also illustrated in Fig.~\ref{fig:post_sn}.
The derived parameters $H_0$, $S_8$, and $\Omega_m$ from all three combined datasets closely match the values reported by Planck under the $\Lambda$CDM assumption~\cite{Planck:2018vyg}, albeit with slightly larger uncertainties due to the increased number of free parameters in the model.
Regarding model comparison, we find that the 4PDE model is statistically favored over $\Lambda$CDM in terms of the minimum chi-square values, with $\Delta \chi^2_{\mathrm{min},\Lambda{\rm CDM}} < 0$ for all three combinations—most notably for CMB+DESY5, but also for CMB+PantheonPlus and CMB+Union3. However, the Bayesian evidence penalizes the model due to its higher complexity, and as a result, $\Lambda$CDM is preferred for all three datasets. This preference is strongest for the CMB+PantheonPlus combination, with $\Delta \ln \mathcal{Z}_{\Lambda{\rm CDM}} = -4.75$, compared to $\Delta \ln \mathcal{Z}_{\Lambda{\rm CDM}} = -1.92$ for CMB+DESY5 and $\Delta \ln \mathcal{Z}_{\Lambda{\rm CDM}} = -2.52$ for CMB+Union3.

We now consider the dataset CMB+DESI and its combinations with three distinct SNeIa compilations (see also Fig.~\ref{fig:post_desi_sn}): CMB+DESI+DESY5, CMB+DESI+PantheonPlus, and CMB+DESI+Union3. 
For CMB+DESI alone, we observe that the mean value of $w_0$ is significantly shifted away from $-1$ ($w_0 = -0.57 \pm 0.24$ at 68\% CL), and it remains in the quintessence regime at approximately $1.8\sigma$. The DE EoS at early times remains in the phantom regime, quantified by the parameter $w_m$, which takes the value $w_m = -2.41^{+0.92}_{-0.52}$ at 68\% CL—consistent with a phantom behavior at more than $1\sigma$.
Regarding the other two free parameters, we find that $a_t$ remains unconstrained, similar to what we observed in the earlier CMB+SNeIa analyses, while $\Delta_{\rm de}$ receives a lower bound: $\log_{10}(\Delta_{\rm de}) > -1.03$ at 95\% CL.
It is worth noting that recent DESI results~\cite{DESI:2024aqx,DESI:2025fii,Ozulker:2025ehg} indicate a preference for phantom crossing around $a \sim 0.7$. However, in our case, since the steepness of the transition ($\Delta_{\rm de}$) is treated as a free parameter, this flexibility broadens the posterior distribution of $a_t$, rendering it unconstrained.
Turning to the derived parameters, we find that CMB+DESI yields a relatively low
value of the Hubble constant,
$H_0 = 64.8^{+2.1}_{-2.4}$ km/s/Mpc at 68\% CL, compared to the Planck baseline
result assuming $\Lambda$CDM~\cite{Planck:2018vyg}.
This notably low value of $H_0$ is driven by the behavior of $w_0$.
The CMB+DESI combination pushes $w_0$ significantly away from $-1$, and, due to
the well-known geometric degeneracy between $w_0$ and $H_0$, which arises from the
fact that the CMB tightly constrains the angular diameter distance to the last
scattering surface, this shift in $w_0$ translates into a lower inferred value of
$H_0$ (see Fig.~\ref{fig:post_desi_sn}).
Consequently, owing to the geometric degeneracy between $H_0$ and the
matter density parameter $\Omega_m$, which originates from the fact that the CMB
primarily constrains the combination $\Omega_m h^2$ ($=\Omega_c h^2+\Omega_b h^2$) rather than $\Omega_m$ and
$H_0$ separately, $\Omega_m$ assumes a higher value than the Planck baseline,
with $\Omega_m = 0.341 \pm 0.024$ at 68\% CL.
From the model comparison perspective, the $\chi^2$ analysis shows that CMB+DESI favors the 4PDE model over $\Lambda$CDM, with $\Delta \chi^2_{\mathrm{min},\Lambda{\rm CDM}} = -9.96$. However, the Bayesian evidence penalizes the extra model complexity, and the comparison instead favors $\Lambda$CDM, albeit weakly, according to the revised Jeffreys’ scale ($\Delta \ln \mathcal{Z}_{\Lambda{\rm CDM}} = -1.25$).  

Now, when DESY5 is combined with CMB+DESI (i.e., for the combined dataset CMB+DESI+DESY5), we find that the mean value of $w_0$ shifts closer to $-1$ ($w_0 = -0.804 \pm 0.066$ at 68\% CL), while its deviation from the cosmological constant increases to approximately $2.97\sigma$, strengthening its quintessence-like nature. The parameter $w_m$ remains in the phantom regime at roughly $1\sigma$ significance. Among the remaining two parameters, $\Delta_{\rm de}$ receives a lower bound ($\log_{10} \Delta_{\rm de} > -0.896$ at 95\% CL), but $a_t$ continues to remain unconstrained.
The derived parameters $H_0$ and $S_8$ are consistent with the Planck baseline values under $\Lambda$CDM~\cite{Planck:2018vyg}. Notably, the inclusion of DESY5 with CMB+DESI significantly reduces the uncertainties in the parameter space compared to the CMB+DESI case alone.
A particularly interesting result in this case is that both the $\chi^2$ analysis and the Bayesian evidence support the 4PDE model over $\Lambda$CDM. In particular, we find a substantial improvement in the fit: $\Delta \chi^2_{\mathrm{min},\Lambda{\rm CDM}} = -18.54$, with a positive Bayes factor of $\Delta \ln \mathcal{Z}_{\Lambda{\rm CDM}} = 2.57$, indicating moderate evidence in favor of the 4PDE model.
In Fig.~\ref{fig:lnZ_chi2}, a graphical comparison between $\Delta \chi^2_{\mathrm{min},\Lambda{\rm CDM}}$ and $\Delta \ln \mathcal{Z}_{\Lambda{\rm CDM}}$ clearly illustrates the preference for the 4PDE model in the CMB+DESI+DESY5 dataset. A mild preference is also observed for the CMB+DESI+Union3 combination, which will be discussed in the following paragraph.

The next two combined datasets, CMB+DESI+PantheonPlus and CMB+DESI+Union3, are particularly interesting because, among all dataset combinations explored in this work, these are the only cases where $\Delta_{\rm de}$ becomes constrained. Specifically, we find $\log_{10} \Delta_{\rm de} = -0.63 \pm 0.70$ (68\% CL, CMB+DESI+PantheonPlus) and $\log_{10} \Delta_{\rm de} = -0.15^{+0.49}_{-0.71}$ (68\% CL, CMB+DESI+Union3). However, in both cases, the parameter $a_t$ remains unconstrained.
The present-day value of the DE equation of state, $w_0$, is pushed closer to $-1$ compared to CMB+DESI alone ($w_0 = -0.57 \pm 0.24$ at 68\% CL). In particular, we obtain $w_0 = -0.899^{+0.050}_{-0.062}$ (CMB+DESI+PantheonPlus) and $w_0 = -0.74 \pm 0.10$ (CMB+DESI+Union3), both at 68\% CL. This places $w_0$ firmly in the quintessence regime ($w > -1$), corresponding to a
deviation from the phantom divide ($w_{\rm de} = -1$) at approximately $2\sigma$ for
CMB+DESI+PantheonPlus and $2.6\sigma$ for CMB+DESI+Union3. The early-time EoS parameter $w_m$ retains a phantom-like behavior, with its statistical significance varying across the datasets.
The derived parameters, such as $H_0$ and $\Omega_m$, show only minor deviations from Planck (under $\Lambda$CDM)~\cite{Planck:2018vyg}. For CMB+DESI+Union3, we find $H_0 = 66.27 \pm 0.87$ km/s/Mpc and $\Omega_m = 0.3242 \pm 0.0088$ at 68\% CL, indicating a mild shift. In contrast, the values returned by CMB+DESI+PantheonPlus are fully consistent with those of Planck within $\Lambda$CDM.
Finally, regarding model comparison statistics, the $\chi^2$ analysis shows a preference for the 4PDE model over $\Lambda$CDM in both combinations ($\Delta \chi^2_{\mathrm{min},\Lambda{\rm CDM}} < 0$). However, the Bayesian evidence provides a more nuanced picture: only CMB+DESI+Union3 very mildly favors the 4PDE model, with $\Delta \ln \mathcal{Z}_{\Lambda{\rm CDM}} = 0.67$, which is classified as \textit{inconclusive} under the revised Jeffreys’ scale. On the other hand, CMB+DESI+PantheonPlus shows a preference for $\Lambda$CDM, with $\Delta \ln \mathcal{Z}_{\Lambda{\rm CDM}} = -1.45$.
Taken together, these results present a mixed picture in terms of model preference, depending on the dataset combination and statistical criterion employed. 

Before concluding this section, we investigate additional cosmological quantities that provide further insight into the behavior of the 4PDE model. In Fig.~\ref{fig:Reconstruction}, we show the evolution of the normalized dark energy density, $\rho_{\mathrm{de}}(a)/\rho_{\mathrm{de}}(a_0)$ (left panel), and the dark energy equation of state $w_{\rm de}(a)$ (right panel), using MCMC samples from the CMB+DESI, CMB+DESI+DESY5, and CMB+DESI+PantheonPlus datasets. The evolution of $\rho_{\rm de}(a)$ reveals an emergent behavior: it grows from early times, reaches a maximum, and then begins to decline. However, we note that this reconstructed evolution is comparable to that allowed
by the CPL parametrization.
Therefore, in terms of reconstructing the dark energy evolution history, the
present 4PDE analysis does not reveal any distinctive features beyond those
already captured by standard two-parameter models. In contrast, the equation of state $w_{\rm de}(a)$ shows a transition from a phantom regime in the past to a quintessence regime at late times, crossing the phantom divide line, as also seen in recent analyses~\cite{Ozulker:2025ehg,Cheng:2025lod}.
This phantom-to-quintessence transition is reminiscent of the behavior found in the CPL parametrization, particularly in DESI analyses~\cite{DESI:2025zgx}. Similar features have been observed in other two-parameter models of dark energy, as discussed in~\cite{Giare:2024gpk}, suggesting that the dynamical behavior captured by our four-parameter model aligns well with recent trends in observational cosmology.
In Fig.~\ref{fig:comoving_H}, we present the best-fit evolution of the comoving Hubble parameter $H(z)/(1+z)$, including $1\sigma$ and $2\sigma$ confidence bands, for the 4PDE model. These results are compared against the DESI DR2 data for $D_H/r_d$. We focus on the CMB+DESI+DESY5 combination, which offers the strongest statistical support for the 4PDE model among the datasets considered. As shown in the figure, the 4PDE model provides an excellent match to the DESI data, especially at high redshift.
Finally, in Fig.~\ref{fig:bao_D}, we compare the predictions of the 4PDE model and $\Lambda$CDM for three different BAO distance measurements from DESI, again using the CMB+DESI+DESY5 dataset. The bottom panel of Fig.~\ref{fig:bao_D} shows the normalized residuals between theoretical predictions and observational data: circles represent $\Lambda$CDM, and crosses correspond to the 4PDE model. As evident from the residuals, the 4PDE model is generally favored across most redshift bins, consistent with the trend observed in Fig.~\ref{fig:lnZ_chi2}, where this dataset combination shows the strongest preference for the 4PDE scenario.

\section{Summary and Conclusions}
\label{sec-summary}

This article investigates a four-parameter dynamical dark energy (4PDE) model aimed at understanding the evolution of dark energy from early times to the present. The free parameters of the model are: $w_0$ (the present-day value of the dark energy equation of state, EoS), $w_m$ (its early-time value), $a_t$ (the scale factor at which the transition from $w_m$ to $w_0$ occurs), and $\Delta_{\rm de}$ (the steepness of the transition). This parametrization was originally introduced by Corasaniti and Copeland~\cite{Corasaniti:2002vg}, and more recently revisited in~\cite{Sharma:2022ifr}. However, unlike other widely studied dynamical dark energy (DDE) models, this and similar four-parameter extensions have not received much attention in the literature. A likely reason is the large number of free parameters, which can lead to degeneracies and weaker constraints. 
Nonetheless, two important considerations motivate renewed attention to such models. First, the true nature of dark energy remains unknown, and a range of observational studies, both model-dependent and non-parametric, support the possibility of dynamical behavior. While the CPL parametrization remains a popular benchmark, the broader search for alternative DDE models is ongoing. Second, the sensitivity and redshift coverage of cosmological datasets are rapidly improving, and previously unconstrained or degenerate parameters may become distinguishable with new data.

In this work, we constrain the 4PDE model using the latest cosmological datasets, including CMB (Planck 2018), BAO from DESI DR2, and three supernova compilations: PantheonPlus, DESY5, and Union3. The results are summarized in Table~\ref{table-4-parameter-DDE}, with graphical representations provided in Figs.~\ref{fig:post_sn},~\ref{fig:post_desi_sn},~\ref{fig:lnZ_chi2},~\ref{fig:Reconstruction},~\ref{fig:comoving_H}, and~\ref{fig:bao_D}.  We find that constraining all four parameters remains challenging:
$a_t$ is not constrained by any dataset, while the constraints on $w_m$ and
$\Delta_{\rm de}$ remain weak.
Only the present-day value of the dark energy equation of state, $w_0$, is well
constrained across all datasets.
This indicates that current cosmological probes do not yet have the sensitivity
required to simultaneously constrain the full dynamics of this four-parameter
extension. Our results consistently indicate a quintessential nature for dark energy at present, with a phantom-like behavior in the past.

One of the key outcomes of this analysis is that, despite the increased number of
free parameters compared to $\Lambda$CDM, the 4PDE model is statistically favored—according to both $\Delta \chi^2$ and Bayesian evidence—for the CMB+DESI+DESY5 combination, which shows moderate evidence in its favor.\footnote{The gain in evidence in favor of the 4PDE model due to
DESY5 is not surprising, as the last two DESI data releases have shown that
evidence for dynamical dark energy is more pronounced in the presence of DESY5
than for the other two SNeIa datasets~\cite{DESI:2024mwx,DESI:2025zgx}.
Although DESI also reported evidence for dynamical dark energy (assuming the CPL
parametrization) when using the remaining two SNeIa datasets, however, one cannot disregard the fact that 
such evidence depends on the underlying dark energy model as well.}
In contrast, the analysis yields inconclusive evidence for the
CMB+DESI+Union3 combination.
For the remaining dataset combinations, the standard $\Lambda$CDM model remains
preferred. In addition to the comparison with $\Lambda$CDM, we assess the statistical
performance of the 4PDE model against the CPL parametrization in
Table~\ref{table-4-parameter-DDE}. 
For the majority of dataset combinations, the two models are statistically
indistinguishable.
The differences in both the goodness of fit and the Bayesian evidence are
negligible, with $|\Delta \chi^2_{\mathrm{min},\mathrm{CPL}}| < 1$ and
$|\Delta \ln \mathcal{Z}_{\mathrm{CPL}}| < 1$.
Although the 4PDE model occupies a significantly larger prior volume due to its
additional free parameters, it does not incur the expected Bayesian penalty.
This indicates that the extra parameters, in particular
$\log_{10}(a_{\rm t})$ and $\log_{10}(\Delta_{\mathrm{de}})$, remain largely
unconstrained by these datasets.
When the data are unable to significantly reduce the allowed parameter space,
the posterior volume remains comparable to the prior volume
($V_{\text{posterior}} \approx V_{\text{prior}}$), effectively canceling the
Occam’s razor penalty typically associated with increased model complexity.
A notable exception is the CMB+DESI+PantheonPlus combination, which favors the
4PDE model with an improvement of
$\Delta \chi^2_{\mathrm{min},\mathrm{CPL}} = -3.74$ and a positive evidence gain
of $\Delta \ln \mathcal{Z}_{\mathrm{CPL}} = 1.14$.
This preference arises because this specific dataset combination provides
tighter constraints on $\log_{10}(\Delta_{\mathrm{de}})$, allowing the model to
effectively exploit its extended parameter freedom to achieve a better fit to
the data. In conclusion, while one dataset combination
(CMB+DESI+PantheonPlus) hints at a weak preference for the 4PDE scenario, current
observational data are generally insufficient to distinguish it from the simpler CPL model.
This suggests that higher-precision data will be required to justify the inclusion of additional degrees of freedom in the dark energy sector. 
These findings suggest that four-parameter DDE models deserve further exploration, particularly with upcoming high-precision cosmological observations. Although one parameter ($a_t$) remains unconstrained in the current analysis, it remains to be investigated whether this is due to limitations of current data or inherent features of the model itself. The preference for this extended model in some dataset combinations provides a compelling motivation to revisit richer DDE parameterizations in future work. 

\begin{acknowledgments}
We thank the referee for many insightful comments which resulted in an improved version of the manuscript. HC acknowledges the support of the China Scholarship Council (CSC) program (Project ID No.\ 202406230341). HC also acknowledges the support by the National Natural Science Foundation of China (NSFC) through the grant No.\ 12350610240 ``Astrophysical Axion Laboratories''. SP acknowledges the partial support by the Department of Science and Technology (DST), Govt. of India under the Scheme ``Fund for Improvement of S\&T Infrastructure (FIST)'' (File No. SR/FST/MS-I/2019/41).  EDV is supported by a Royal Society Dorothy Hodgkin Research Fellowship. 
The authors acknowledge the use of High-Performance Computing resources from the IT Services at the University of Sheffield.
This article is based upon work from COST Action CA21136 Addressing observational tensions in cosmology with systematics and fundamental physics (CosmoVerse) supported by COST (European Cooperation in Science and Technology). 
\end{acknowledgments}
\bibliography{biblio}
\end{document}